\documentclass[twoside]{elsart}
\usepackage{epsfig}

\columnwidth\textwidth

\begin{document}

\begin{frontmatter}
\title{Comparison of AGASA data with\\ CORSIKA simulation}
                       
\author[fukui]{M. Nagano\thanksref{corresponding}}
\author[fzk]{D. Heck}
\author[saitama]{K. Shinozaki}
\author[saitama]{N. Inoue}
\author[leeds]{J. Knapp}

\address[fukui]{Department of Applied Physics and 
Chemistry, Fukui University of Technology, Fukui, 910-8505 Japan}
\address[fzk]{Institut f\"ur Kernphysik III, Forschungszentrum Karlsruhe,\\
Postfach 3640, D 76021 Karlsruhe, Germany} 
\address[saitama]{Department of Physics, Saitama University, Urawa 338-8570, Japan}
\address[leeds]{Department of Physics and Astronomy, University of Leeds,\\
Leeds LS2 9JT, United Kingdom} 

\thanks[corresponding]{E-mail: mnagano@ccmails.fukui-ut.ac.jp}

\begin{abstract}
An interpretation of AGASA (Akeno Giant Air Shower Array) data by
comparing the experimental results with the simulated ones by CORSIKA
(COsmic Ray SImulation for KASCADE) has been made.  General features of
the electromagnetic component and low energy muons observed by AGASA can
be well reproduced by CORSIKA.  The form of the lateral distribution of
charged particles agrees well with the experimental one between a few
hundred metres and 2000 m from the core, irrespective of the hadronic
interaction model studied and the primary composition (proton or iron).
It does not depend on the primary energy between 10$^{17.5}$ and
10$^{20}$ eV as the experiment shows.  If we evaluate the particle
density measured by scintillators of 5 cm thickness at 600 m from the
core ($S_0$(600), suffix 0 denotes the vertically incident shower) by
taking into account the similar conditions as in the experiment, the
conversion relation from $S_0$(600) to the primary energy is expressed
as $E [\mbox{eV}] = 2.15 \times 10^{17} S_0(600)^{1.015},$
within 10\% uncertainty among the models and composition used, which
suggests the present AGASA conversion factor is the lower limit.  Though
the form of the muon lateral distribution fits well to the experiment
within 1000 m from the core, the absolute values change with hadronic
interaction model and primary composition.  The slope of the
$\rho_{\mu}$(600) (muon density above 1~GeV at 600 m from the core)
vs. $S_0$(600) relation in experiment is flatter than that in simulation
of any hadronic model and primary composition.  Since the experimental
slope is constant from $10^{15}$ eV to $10^{19}$ eV, we need to study
this relation in a wide primary energy range to infer the rate of change
of chemical composition with energy.
\end{abstract}

\end{frontmatter}

\section{Introduction}
After the observation of cosmic rays with energies greatly exceeding the
Greisen-Zatsepin-Kuzmin (GZK) cutoff energy \cite{grei66} by the Fly's
Eye \cite{bird95} and the Akeno Giant Air Shower Array (AGASA)
\cite{hayashida94}, seven events exceeding 10$^{20}$ eV have been
reported by the AGASA group \cite{takeda98}.

To estimate the primary energy of giant air showers observed by the
AGASA, the particle density at a distance of 600 m from the shower axis
($S_0$(600), suffix 0 denotes the vertically incident shower) is used,
which is known to be a good energy estimator \cite{hillas}.  The
conversion factor from $S_0$(600) [m$^{-2}$] to primary energy ($E_0$
[eV]) at Akeno level is derived by simulation \cite{dai88} based on the
COSMOS program by Kasahara {\it et al.} \cite{kasahara} with the 
QCDjet\footnote{The QCDjet model is based on QCD and contains a mild
scaling violation in the fragmentation region, and a large one in the
central region.  It includes also production of hard mini-jets.  This
model is extensively used for interpretation of the emulsion chamber
experiments at Mt.Fuji, Mt.  Kambala (China) and Mt. Yanbajin (Tibet)
and shows good agreement with the experiments in the fragmentation
region, where reliable data can not be obtained from accelerator
experiments.} model \cite{ding} and the relation
\begin{equation}
 E_0 [\mbox{eV}] = 2.0 \times 10^{17}  S_0(600)^{1.0}
\label{eq-energy1}
\end{equation}
is used.  This relation holds within 20\%, independent of primary
composition and interaction models used \cite{dai88}.  The deviation of
the energy spectrum determined in this way from that determined by the 1
km$^2$ array at Akeno is about 10\% higher in energy around 10$^{18}$ eV
as shown in \cite{nagano84}.  That is,
\begin{equation}
 E_0 [\mbox{eV}] = 1.8 \times 10^{17} S_0(600)^{1.0}
\end{equation}
must be applied instead of Eq. \ref{eq-energy1} to be in agreement with
that from the lower energy region.

\begin{figure}
\begin{center}
\epsfig{file=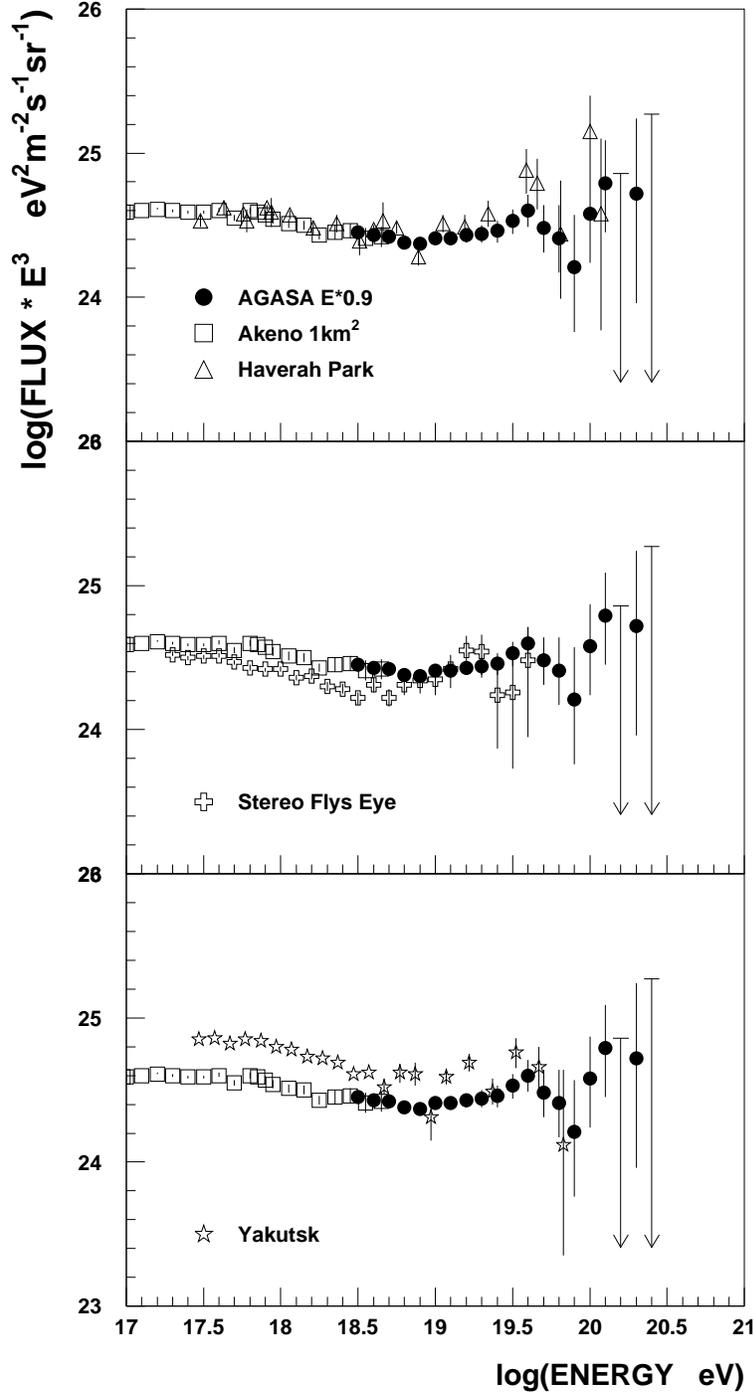,width=10cm,bbllx=1,bburx=346,bblly=14,bbury=660,clip=}
\end{center}
\caption{\small Energy spectra from the Haverah Park, Yakutsk, stereo Fly's
Eye experiments are compared with that from the AGASA experiments.}
\label{all_spectrum}
\end{figure}

The AGASA energy spectrum in the highest energy region, which is
adjusted to the one of the 1 km$^2$ array at Akeno\footnote{The spectrum
determined between 10$^{14.5}$ eV and 10$^{19}$ eV using the arrays with
different detector spacing at Akeno fits very well with extrapolation of
those obtained from direct measurement on balloons and satellites, and
with the Tibet result obtained through the observation of the shower at
the height of its maximum development.  Therefore the AGASA spectrum may
be better to normalize to the spectrum by 1 km$^2$ array.}
\cite{nagano84} are compared with the spectra from the Haverah Park
\cite{law91}, the Yakutsk \cite{afa93} and the stereo Fly's Eye
\cite{bird94} experiments in Fig. \ref{all_spectrum}.  All four spectra
agree with each other within $\pm15\%$ in energy.  It should be noted
that the energy assignment in each experiment has been done separately
by each group.  The Fly's Eye experiment measures the longitudinal
development of electrons above the observation level and is
calorimetric.  The Yakutsk group determined the energy conversion
relation experimentally by measuring not only the lateral distributions
of electrons and muons but also the energy loss of electrons above the
observation level from the \v{C}erenkov lateral distribution at the
observation level.  The Haverah Park experiment used water-\v{C}erenkov
detectors and the AGASA uses plastic scintillation detectors, the energy
conversions of both experiments rely on different simulation codes.
However, the energy assignment is in fairly good agreement as shown by
the water tank experiment at Akeno \cite{sakaki}.

On the other hand, Cronin \cite{cronin97} argued that the AGASA energy
spectrum should be about 30\% higher in energy, if the energy is
assigned in each experiment according to the simulation results based on
the MOCCA program with the SIBYLL hadronic interaction model. According
to their simulation, the relation should be
\begin{equation}
 E_0  [\mbox{eV}] = 3.0 \times 10^{17} S_0(600)^{1.0}
\end{equation}
at Akeno level.  That is, the AGASA energy spectrum results in much
higher intensity than other experiments at the same energy.

Recently Kutter pointed out that the difference is due to the large
contribution of low energy photons \cite{kutter98} in the scintillators
in case of AGASA.  Though the energy loss of electrons and photons in
scintillator ($\rho_{sc}$) in units of a vertically traversing muon is
independent of scintillator thickness, the ratio of $\rho_{sc}$ to the
number of charged particles (R$_{sc/ch}$) depends on the core distance.
R$_{sc/ch}$ is about 1.1 within 100 m from the core and agrees with the
experiment \cite{shibata65}. However, it increases as the core distance
increases.  It becomes about 1.2 around 600 m from the core in case of
MOCCA (SIBYLL) and 1.4 in case of CORSIKA (QGSJET) simulations.  The
difference between the simulation code is due to the difference of
energy spectra of low energy photons and electrons far from the
core. This ratio R$_{sc/ch}$ depends also on the number of charged
particles which increases as the cutoff energy of electromagnetic
component decreases.

Therefore, it is important to evaluate the energy conversion factor
Eq. \ref{eq-energy1} used in AGASA experiment with simulation programs
other than the COSMOS simulation, by taking into account the energy
losses of low energy photons and electrons in scintillator.

Since the total number of muons in an extensive air shower with a fixed
primary energy depends on the chemical composition, the possibility of
discriminating heavy particles from protons has been extensively studied
by many authors. For example, Dawson {\it et al}. \cite{dawson98} argues
that the muon density at 600 m from the core, $\rho_{\mu}$(600),
vs. energy relation observed at Akeno can be consistently explained by
the change of composition from heavy around 10$^{17.5}$ eV to light
around 10$^{19}$ eV, which is claimed by the Fly's Eye group
\cite{gaisser93}.  However, they avoided the discussion of change of
composition around 10$^{17.5}$ eV which is a point of disagreement
between the Fly's Eye experiment and the Akeno experiment.  According to
their simulation result \cite{dawson98}, the Fly's Eye data shows 100\%
iron below 10$^{17.5}$ eV and AGASA data shows heavier than iron below
10$^{17}$ eV.  It is important not only to examine the simulation
results with other hadronic interaction models, but also to examine the
experimental relation of $\rho_{\mu}$(600) vs. $S_0$(600) in the lower
energy region to see whether there is any change of slope in the
relation.

As the arrival directions of observed EAS are inclined from the zenith,
it is necessary to convert $S_{\theta}$(600), where $\theta$ represents
the zenith angle, to $S_0$(600).  For zenith angles less than
45$^{\circ}$ ($\sec\theta\sim$1.4),
\begin{equation}
 S_{\theta}(600)=S_0(600) \exp \left[- {X_0\over \Lambda_{att}}(\sec\theta - 1)
  \right]
\label{zenith-dep1}
\end{equation}
is used, where the attenuation length $\Lambda_{att}=500$ g cm$^{-2}$
and $X_0$ is the atmospheric depth at Akeno (920 g cm$^{-2}$)
\cite{nagano92}.  The attenuation of $S_0$(600) can be determined from
integral $S_{\theta}$(600) spectra at various $\theta$ by assuming
$S_{\theta}$(600) at constant intensity in different zenith angles from
primaries of similar energy.  If we use data of zenith angles up to
55$^{\circ}$,
\begin{equation}
 S_{\theta}(600)=S_0(600) \exp \left[- {X_0\over \Lambda_1}(\sec\theta - 1)
                                    - {X_0\over \Lambda_2}(\sec\theta - 1)^2
  \right]
\label{zenith-dep2}
\end{equation}
is better fitted to the data, where $\Lambda_1=\Lambda_{att}=500$ g
cm$^{-2}$ and $\Lambda_2=594$ g cm$^{-2}$. This was derived, in
\cite{yoshida94}, between $2\times10^{18}$ eV and $5\times10^{19}$ eV.
Since this attenuation with zenith angle is related closely to the
properties of hadronic interactions at ultra-high energy and to the
composition of primary cosmic rays, the interpretation of the present
result by simulation result is important.

In this article, we use the EAS simulation program CORSIKA ( COsmic Ray
SImulation for KASCADE) \cite{heck98a} which was developed at Karlsruhe
and is now widely distributed and used by various experiments from TeV
gamma-rays to the highest energy region.  Recently the program was
improved to be used effectively at the highest observed energy
\cite{heck98b}.  By employing the effective thin sampling procedure, the
computing time is considerably reduced and hence various combinations of
simulation conditions with different energies have been realized in this
study.

\section{AGASA data}
 
\subsection{Particle density measured by scintillation detectors on surface} 

The AGASA is the Akeno Giant Air Shower Array covering over 100 km$^2$
area in operation at the village of Akeno about 130 km west of Tokyo, to
study ultra high energy cosmic rays (UHECR) above 10$^{19}$ eV
\cite{chiba92,ohoka96}.

The lateral distribution of electrons and the shower front structure far
from the core, which are important to obtain $S_{\theta}$(600), are
determined with the numerous detectors of 1 km$^2$ array (A1) when
showers hit inside the Akeno Branch.  The density at core distance r is
expressed by the function as
\begin{equation}
\rho=N_eC_eR^{-\alpha}(1+R)^{-(\eta-\alpha)}\left( 1.0+
\left({r\over1000}\right)^2 \right)^{-0.6},
\label{eq-elec}
\end{equation}
where $R=r/R_M$, $C_e$ is a normalization factor and $R_M$ is a
Moli\`{e}re unit (MU) at a height of two radiation lengths above the
Akeno level (91.6 m at Akeno) \cite{yoshida94}.  A fixed value of
$\alpha=1.2$ is used and $\eta$ is expressed as
\begin{equation}
\eta = 3.97 - 1.79 (\sec \theta -1).
\label{eq-eta}
\end{equation}

Recently this function was found to be valid up to 3000 m from the core
and the highest observed energy $\sim10^{20}$ eV \cite{sakaki98}.  It is
important to examine with simulations whether such an energy
independence can be understood with a pure proton composition of the
primary particles.

\subsection{Muons measured by proportional counters under the absorber} 

At AGASA, muons of energies above 0.5 GeV are measured under the
lead/iron or concrete shielding.  At the first stage of AGASA
experiment, the lateral distribution of muons far from the core was
determined with eight muon detectors of 25 m$^2$ each (threshold muon
energy : 1 GeV) in the central part of the Akeno Observatory triggered
by the AGASA scintillation detectors on the surface.  The lateral
distribution of muons above 1 GeV is expressed by the following equation
\cite{hayashida95}.
\begin{equation}
\rho_{\mu}=N_{\mu}\left( \frac{C_{\mu}}{R_o^2} \right)
R^{-0.75}(1+R)^{-2.52}\left( 1.0+ \left( {r\over800} \right)^3
\right)^{-0.6},
\label{eq-muon}
\end{equation}
where $N_{\mu}$ is a total number of muons and $R=r/R_o$. $C_{\mu}$ is a
normalization factor and $R_o$ is a characteristic distance and is
expressed by the following equation as a function of zenith angle
$\theta$,
\begin{equation}
  \log(R_o)=(0.58 \pm 0.04)(\sec\theta - 1) + (2.39 \pm 0.05).
\label{eq-rzero}
\end{equation}
This formula can be applied up to 3000 m from the core for showers above
3$\times10^{19}$ eV as shown in Doi {\it et al}.  \cite{doi95}, though
the experimental error is still larger than $\pm$50\% beyond 1000 m from
the core in the highest observed energy region.  To combine the AGASA
data of threshold energy of 0.5 GeV with Eq. \ref{eq-muon} of 1 GeV
threshold, the AGASA data is reduced by a factor 1.4, which is
determined at 600 m from the core, in the whole distance range
\cite{matsu85}.

\section{CORSIKA} 
In CORSIKA five high energy hadronic interaction models are available
and a comparison of available interaction models has been made up to
10$^{17}$ eV \cite{knapp}.  Among those models QGSJET \cite{kalmykov93},
which includes minijet production in hadronic interactions, seems to be
the best to extrapolate to the highest observed energy range.  For
comparison SIBYLL \cite{fletcher94}, which is based on a QCD minijet
model, is also tried in a part of the analysis.

To see the general aspects of the simulation results in the highest
energy region, simulations with different interaction models (QGSJET,
SIBYLL), primary energies (10$^{17.5}, 10^{18.0}$, 10$^{18.5},
10^{19.0}$, 10$^{19.5}, 10^{20.0}$ eV), primary masses (proton, iron),
thinning levels (10$^{-5}$, 10$^{-6}$), zenith angles 
(0$^{\circ}$,
29.6$^{\circ}$ 
39.7$^{\circ}$, and
51.3$^{\circ}$
or $\sec\theta=1.0$, 1.15, 1.3, and 1.6, respectively), 
and cutoff energies of electromagnetic component
(1.0 MeV, 0.1 MeV, 0.05 MeV) have been performed. Cutoff energies of
hadrons and muons are fixed in this series of simulation at 100 MeV and
10 MeV, respectively.

In each combination, 10 showers are simulated and the average values at
the Akeno level are summarized.  To study fluctuations, 30 showers are
used with a thinning level of 10$^{-6.5}$ for a limited combinations of
conditions.

In the following all simulation results are from CORSIKA, unless
otherwise noted.

\section{Results}

\subsection{Lateral distribution of electrons, photons and muons}

The average lateral distributions of photons, electrons and muons do not
 depend on thinning levels between 10$^{-5}$ and 10$^{-6}$.  However,
 thinning levels lower than 10$^{-6}$ are required to discuss the
 detailed behavior of scintillator response far from the core, taking
 into account the energy spectrum of individual particles.  First the
 lateral distributions of electrons and photons are compared with two
 different cutoff energies of the electromagnetic component, 1.0 MeV and
 0.05 MeV.  Though there are increases in the number of electrons and
 photons with decreasing cutoff energy (about 10$\sim$20\%\footnote{The
 fraction depends on the primary composition and the hadronic
 interaction model used.} for the cutoff energies between 1.0 MeV and
 0.05 MeV), the change of the form\footnote{The densities may be shifted
 vertically to compare the form of the lateral distribution with the
 experimental points, since the primary energy in the experiment is
 given by the conversion equation by Dai {\it et al}.} of the lateral
 distribution of charged particles is not significant.  Therefore the
 cutoff energy of 1.0 MeV is used in the following simulations to save
 computing time and disk space for individual particle information.

\subsubsection{Energy dependence}

\begin{figure}
\begin{center}
\epsfig{file=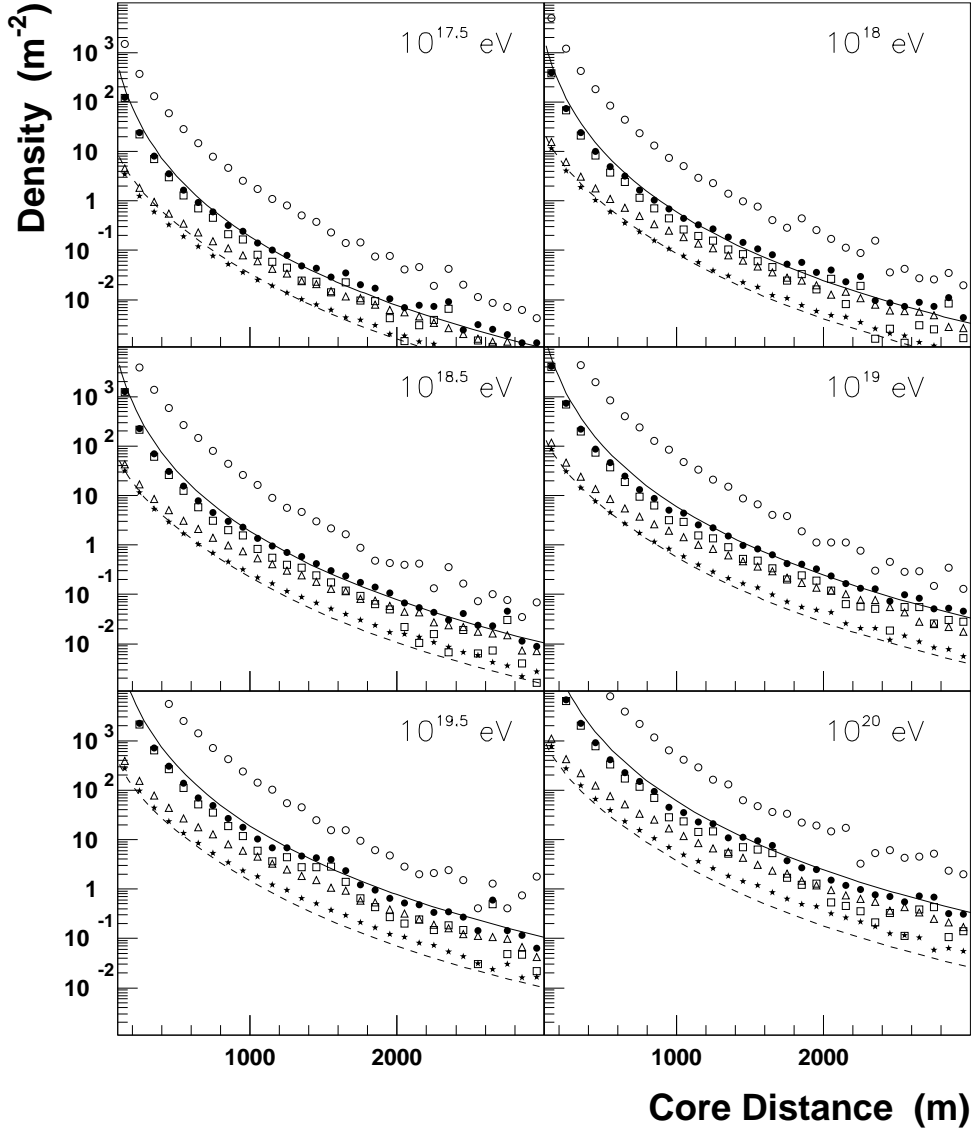,width=13cm,bbllx=7,bburx=423,bblly=14,bbury=500,clip=}
\end{center}
\caption{\small The lateral distributions of photons (open circles),
electrons (open squares) and muons (open triangles) above 10 MeV and
muons (stars) above 1 GeV simulated with the QGSJET model.  Charged
particles which are addition of electrons and muons above 10~MeV are
also plotted by closed circles.  The thinning level is 10$^{-6}$ and the
cutoff energy for muons is 10 MeV and those for photons and electrons
are 1.0 MeV.  Solid and dashed lines are the AGASA empirical formula on
the surface (Eq. \ref{eq-elec}) and under absorber (Eq. \ref{eq-muon}),
respectively.}
\label{phelmu-6}
\end{figure}

In Fig. \ref{phelmu-6}, the primary energy dependence of the lateral
distributions of photons, electrons and muons is shown for proton
primaries with the QGSJET model.  Closed circles are the lateral
distributions of charged particles (muons and electrons) and the solid
and dashed lines represent the empirical formula of the AGASA
experiment.  Considering the difference of the primary energy assignment
in the experiment and the simulation, if we shift the simulated points
upward to fit the experimental ones within 1000 m from the core, the
shapes are in good agreement with experimental results up to 1000 $\sim$
2000 m from the core.

The energy {\em independence} of the lateral distribution of charged
particles at the surface between 10$^{17.5}$ eV and 10$^{20}$ eV
observed by the AGASA is well supported by the QGSJET model.

\subsubsection{Hadronic model dependence}

\begin{figure}
\begin{center}
\epsfig{file=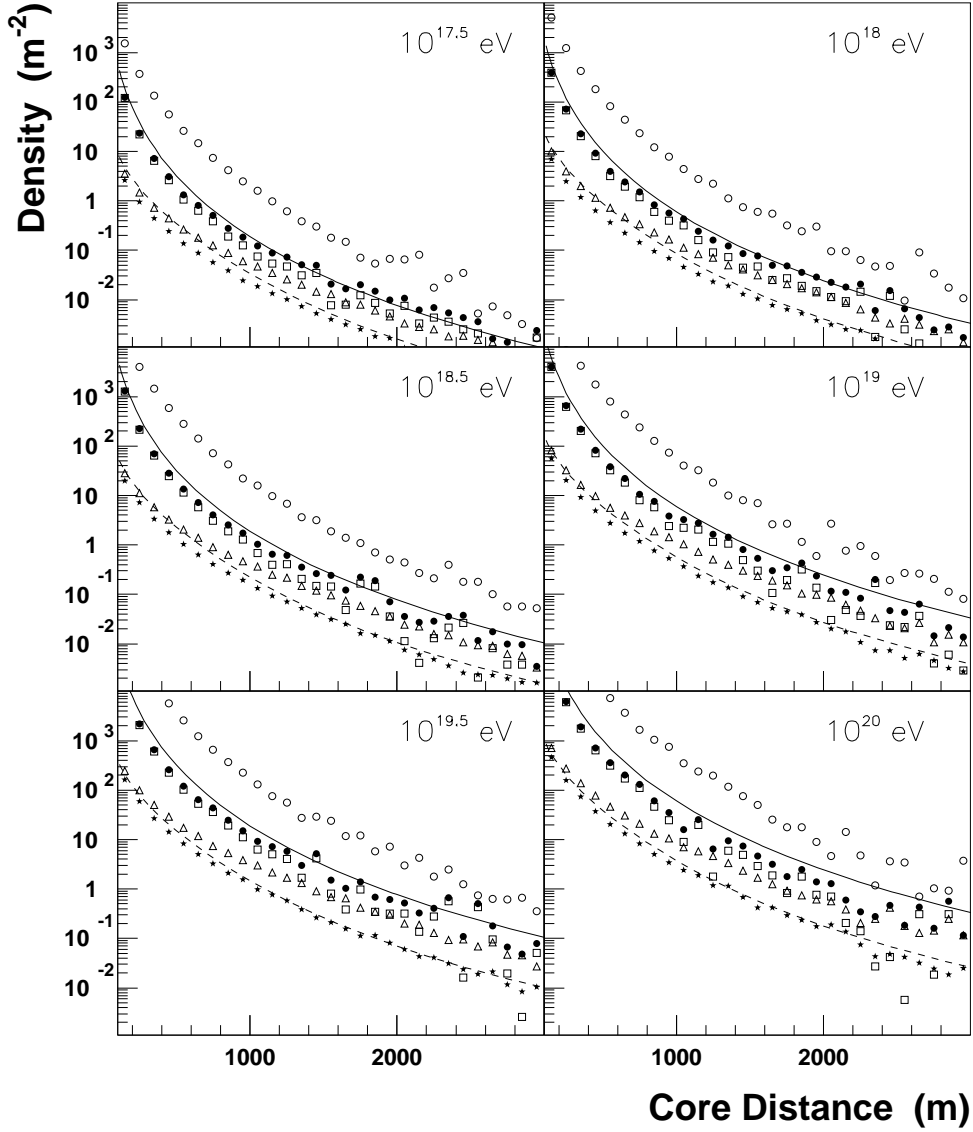,width=13cm,bbllx=7,bburx=423,bblly=14,bbury=500,clip=}
\end{center}
\caption{\small Same as Fig. \ref{phelmu-6} except for the SIBYLL hadronic interaction
model being used.}
\label{fit-p-s}
\end{figure}

The lateral distributions of charged particles from SIBYLL are shown in
Fig. \ref{fit-p-s}.  If we shift the simulated values upward, the
expected lateral distributions are also fitted to the experiment:
however, the deviation from experiment slightly increases as energy
increases (Fig. \ref{fit-p-s}).

While there is no big difference in charged particles for QGSJET and
SIBYLL calculations, SIBYLL underestimates muons systematically by about
a factor 1.6 with respect to QGSJET, and this would lead to a different
interpretation of the experimental showers concerning their mass
composition.  The absolute values are related to the assignment of the
primary energy and will be discussed in Section 5.3.

\subsubsection{Primary composition dependence}

\begin{figure}
\begin{center}
\epsfig{file=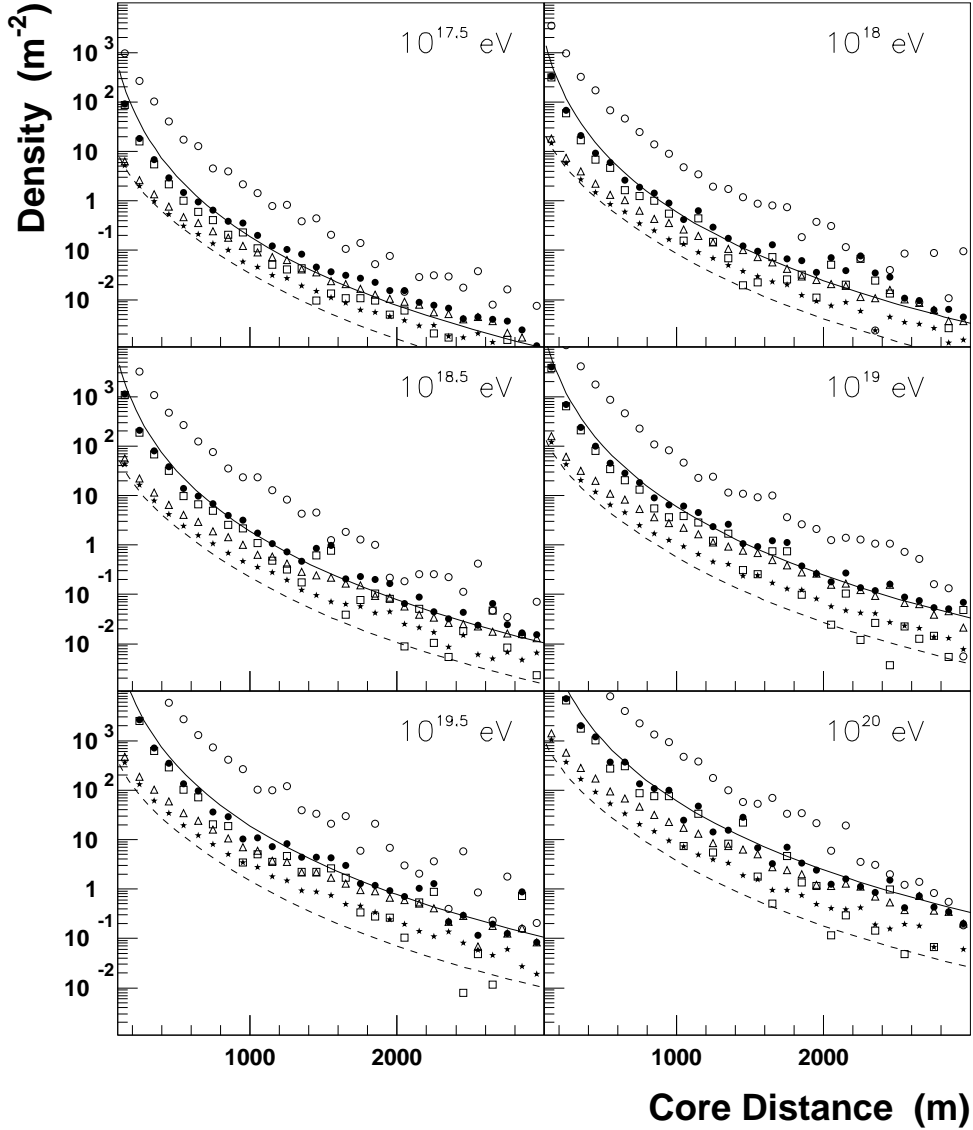,width=13cm,bbllx=7,bburx=423,bblly=14,bbury=500,clip=}
\end{center}
\caption{\small Same as Fig. \ref{phelmu-6}, except for the primary particle being iron.}
\label{fit-fe-q}
\end{figure}

In Fig. \ref{fit-fe-q}, the lateral distributions of charged particles
from iron primary are shown for six different primary energies.  Energy
{\em independence} of the form of the lateral distribution on primary
energy also holds for the iron primary.  Therefore the energy {\em
independence} of lateral distribution of AGASA experiment can be
understood irrespective of primary composition.

The difference of charged particle densities between proton and iron
primaries is about 10\% around 600 m from the core and those of muons
above 1 GeV is about 50\%.

\subsection{Muon lateral distributions with different cutoff energies}

\begin{figure}
\begin{center}
\epsfig{file=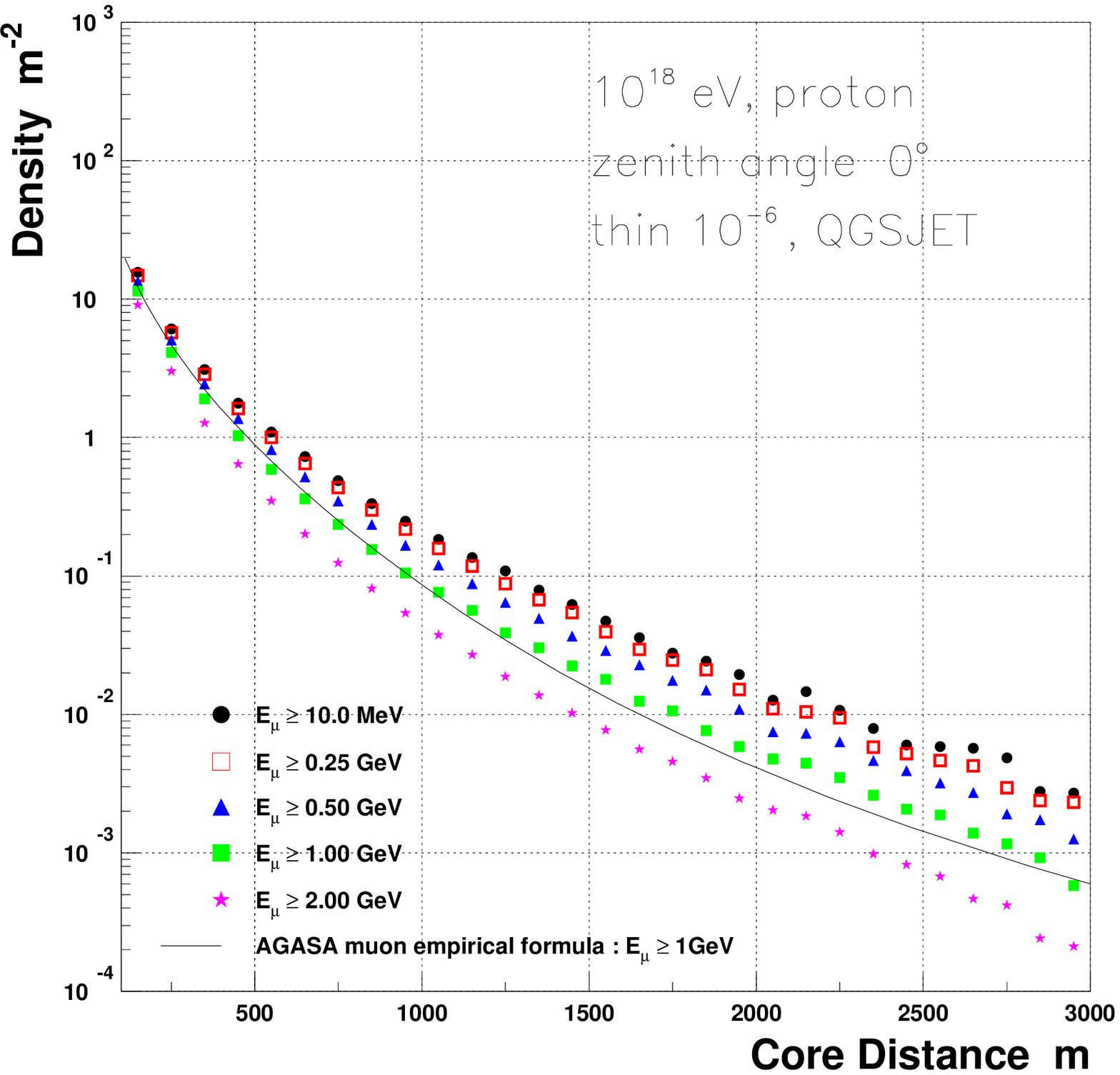,width=7cm,bbllx=0,bburx=522,bblly=18,bbury=520,clip=}
\hfill
\epsfig{file=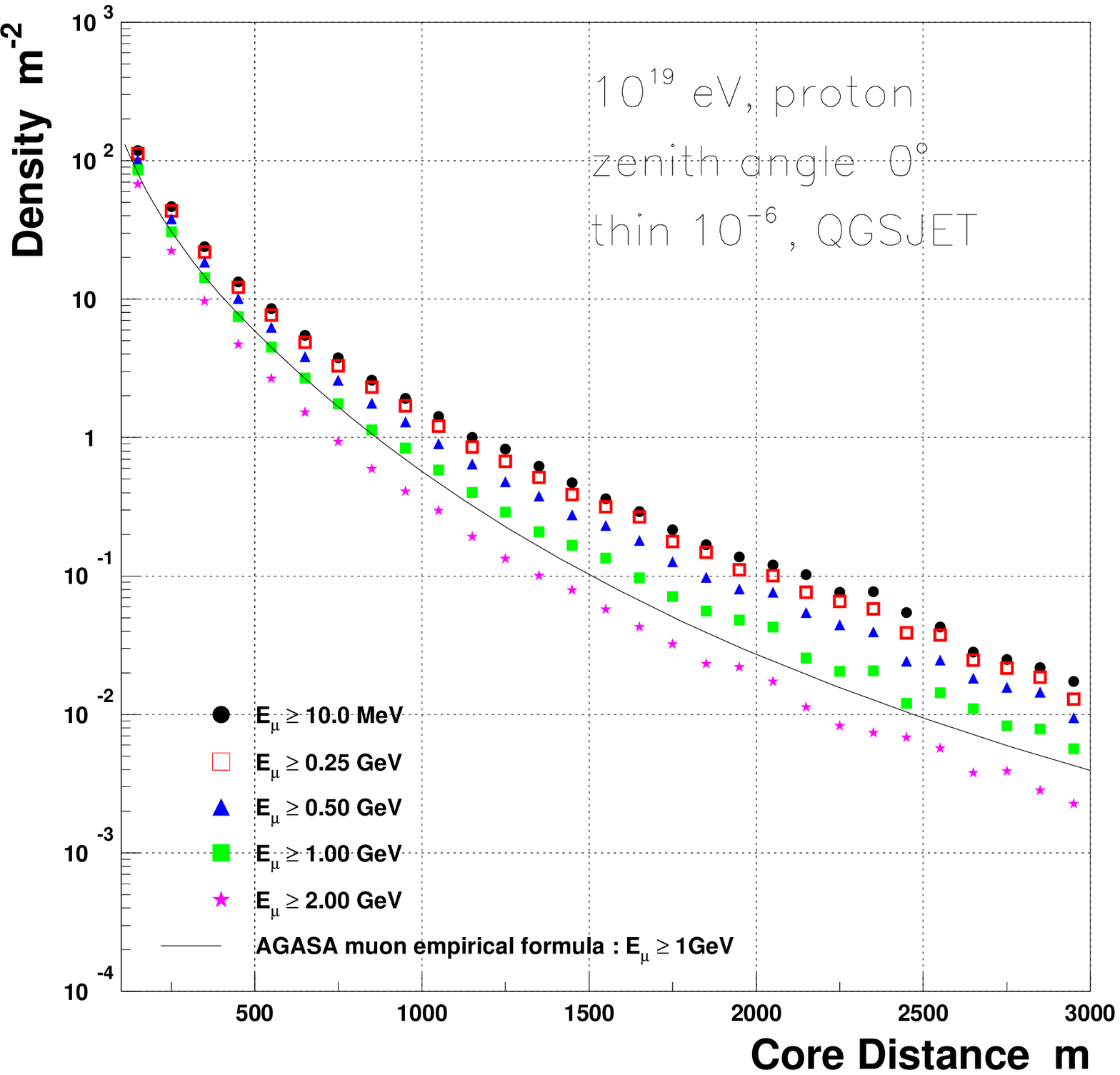,width=7cm,bbllx=0,bburx=522,bblly=18,bbury=520,clip=}
\epsfig{file=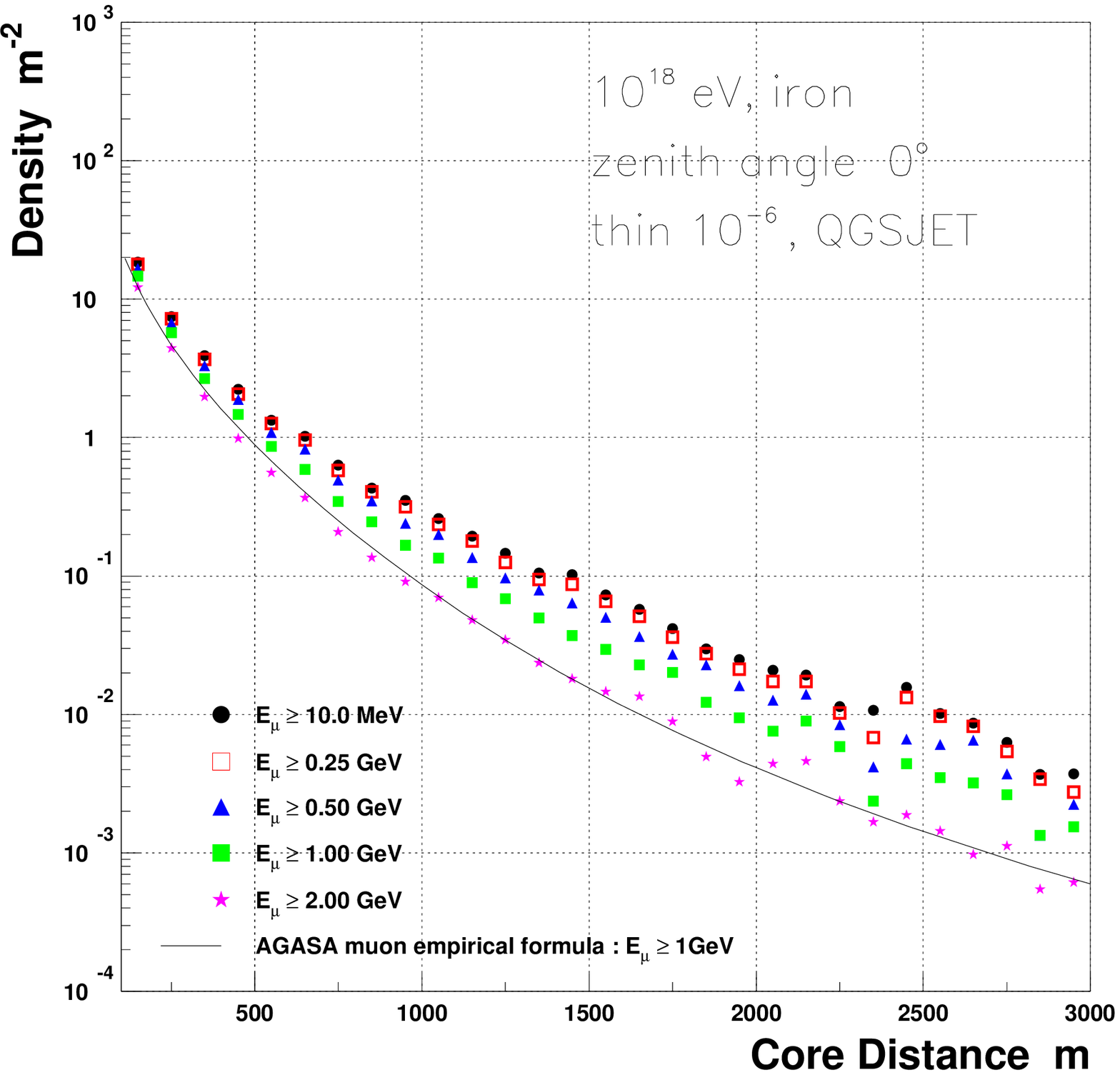,width=7cm,bbllx=0,bburx=522,bblly=18,bbury=520,clip=}
\hfill
\epsfig{file=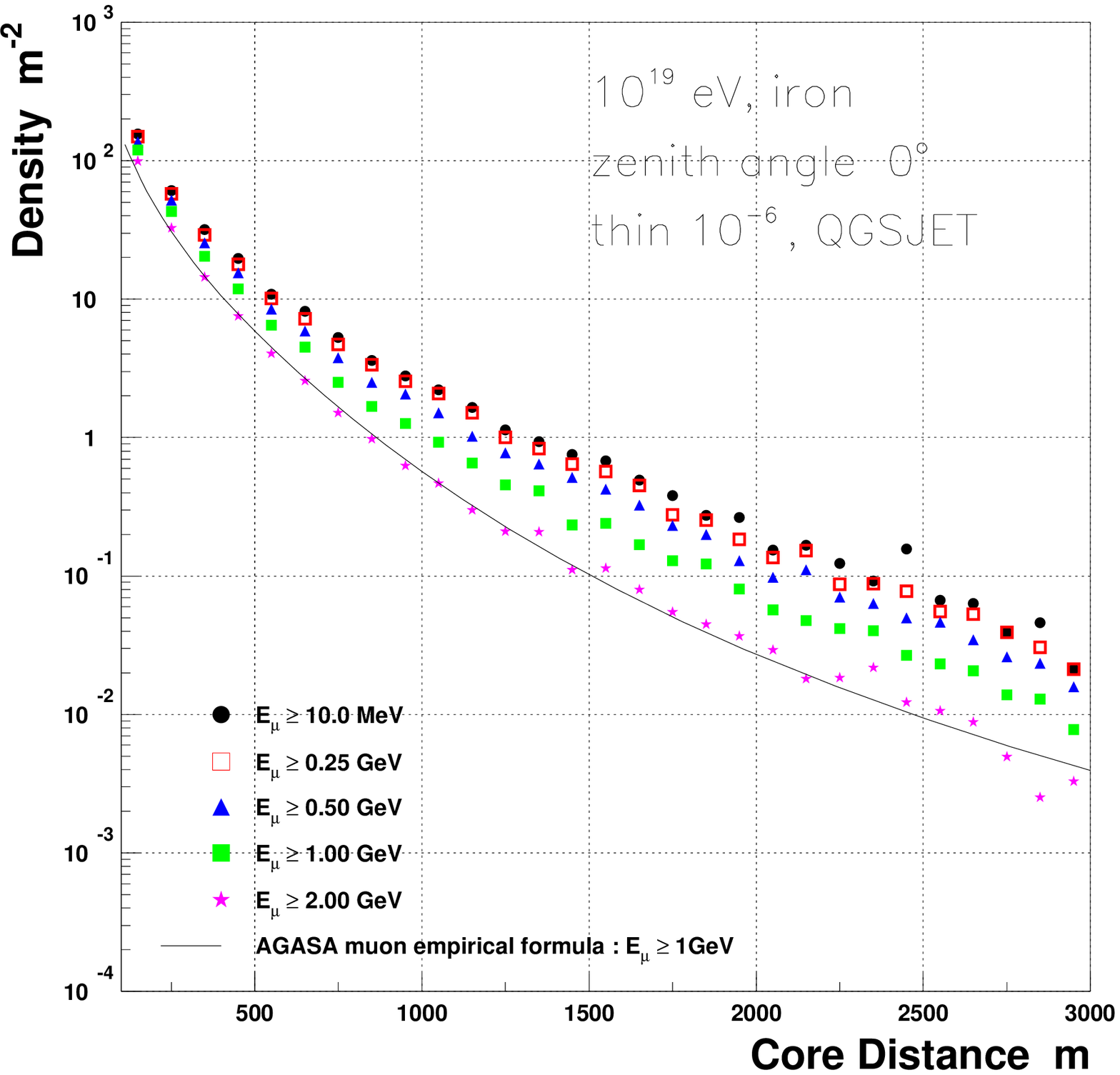,width=7cm,bbllx=0,bburx=522,bblly=18,bbury=520,clip=}
\end{center}
\caption{\small The lateral distributions of muons of cutoff energies above
10 MeV (top), 0.25 GeV, 0.5 GeV, 1.0 GeV and 2.0 GeV (bottom).  QGSJET
model. The thinning level is 10$^{-6}$.  Upper figures are proton
primaries and lower ones are iron.  The incident energies are 10$^{18}$
eV (left) and 10$^{19}$ eV (right), respectively.  Symbols as in Fig. \ref{phelmu-6}.}
\label{mulat1}
\end{figure}

\begin{figure}
\begin{center}
\epsfig{file=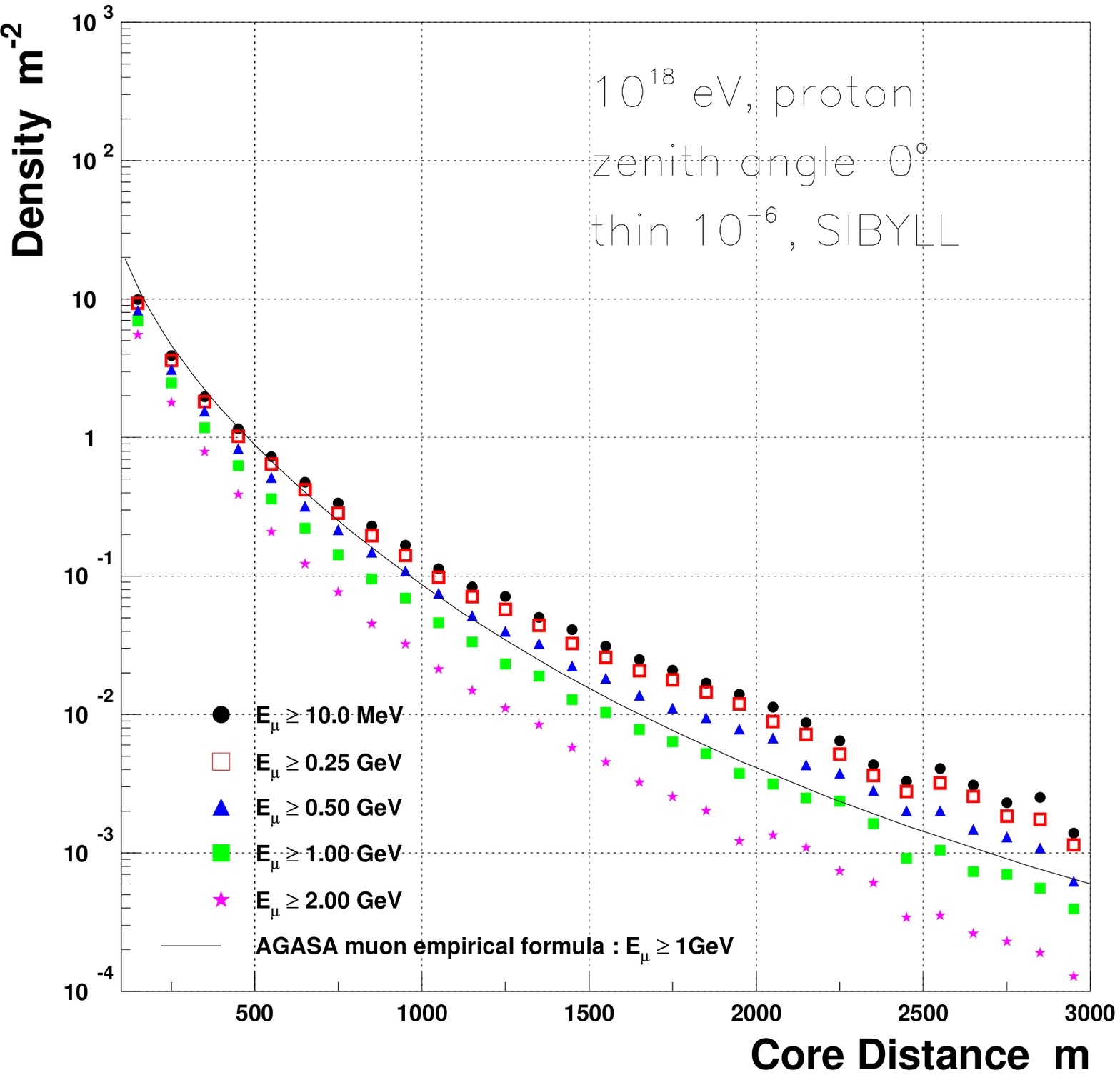,width=7cm,bbllx=0,bburx=522,bblly=18,bbury=520,clip=}
\hfill
\epsfig{file=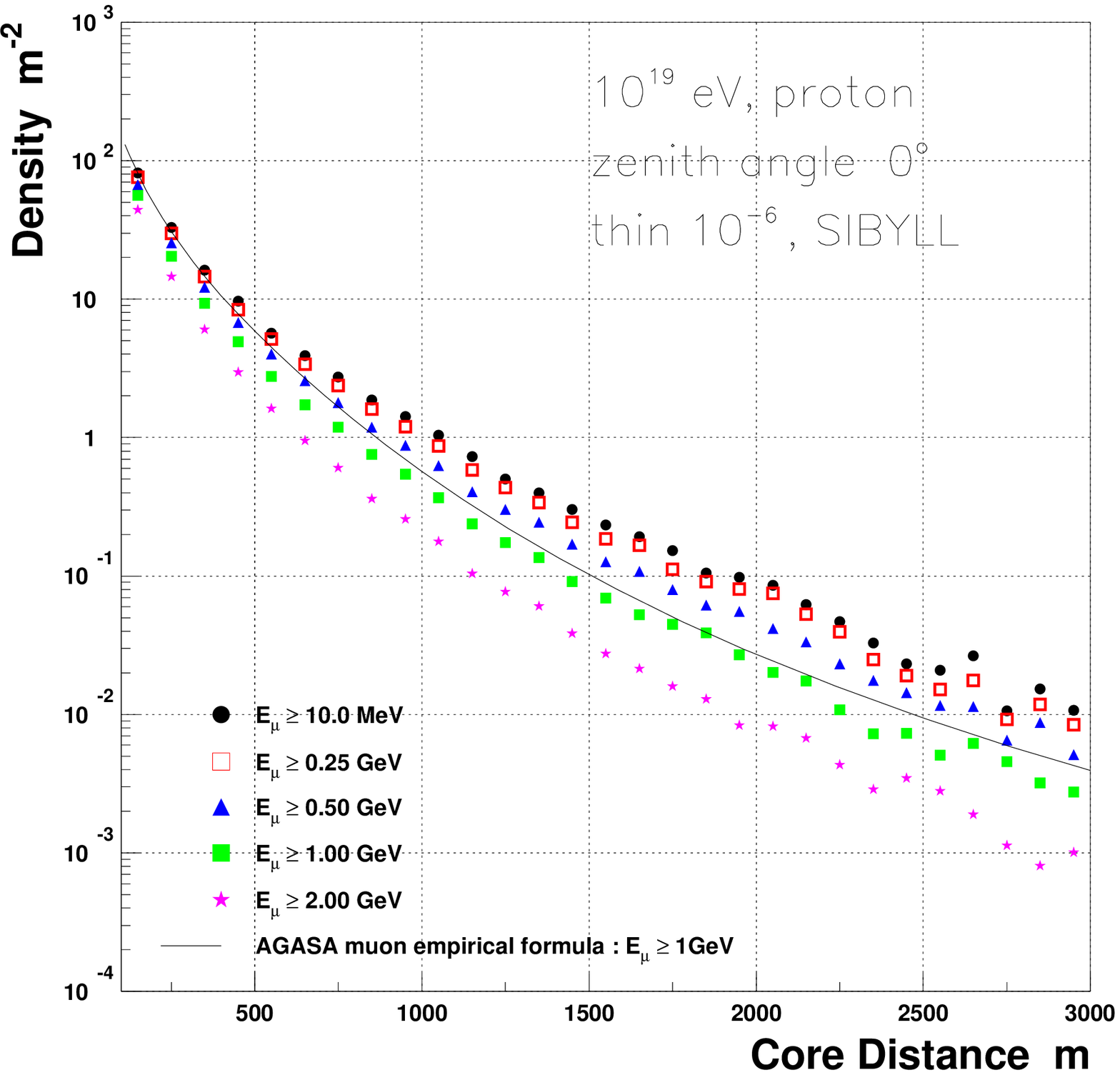,width=7cm,bbllx=0,bburx=522,bblly=18,bbury=520,clip=}
\epsfig{file=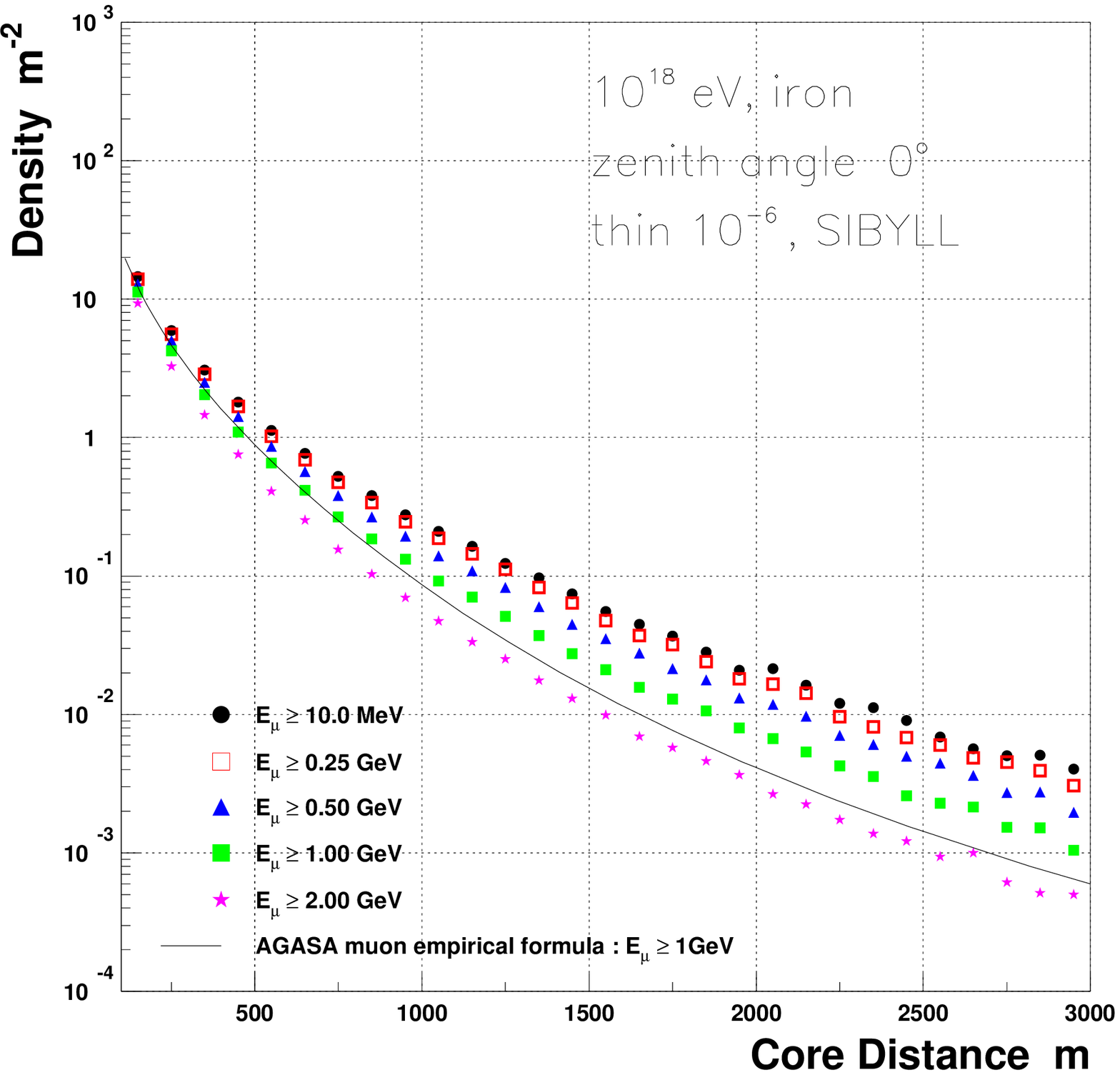,width=7cm,bbllx=0,bburx=522,bblly=18,bbury=520,clip=}
\hfill
\epsfig{file=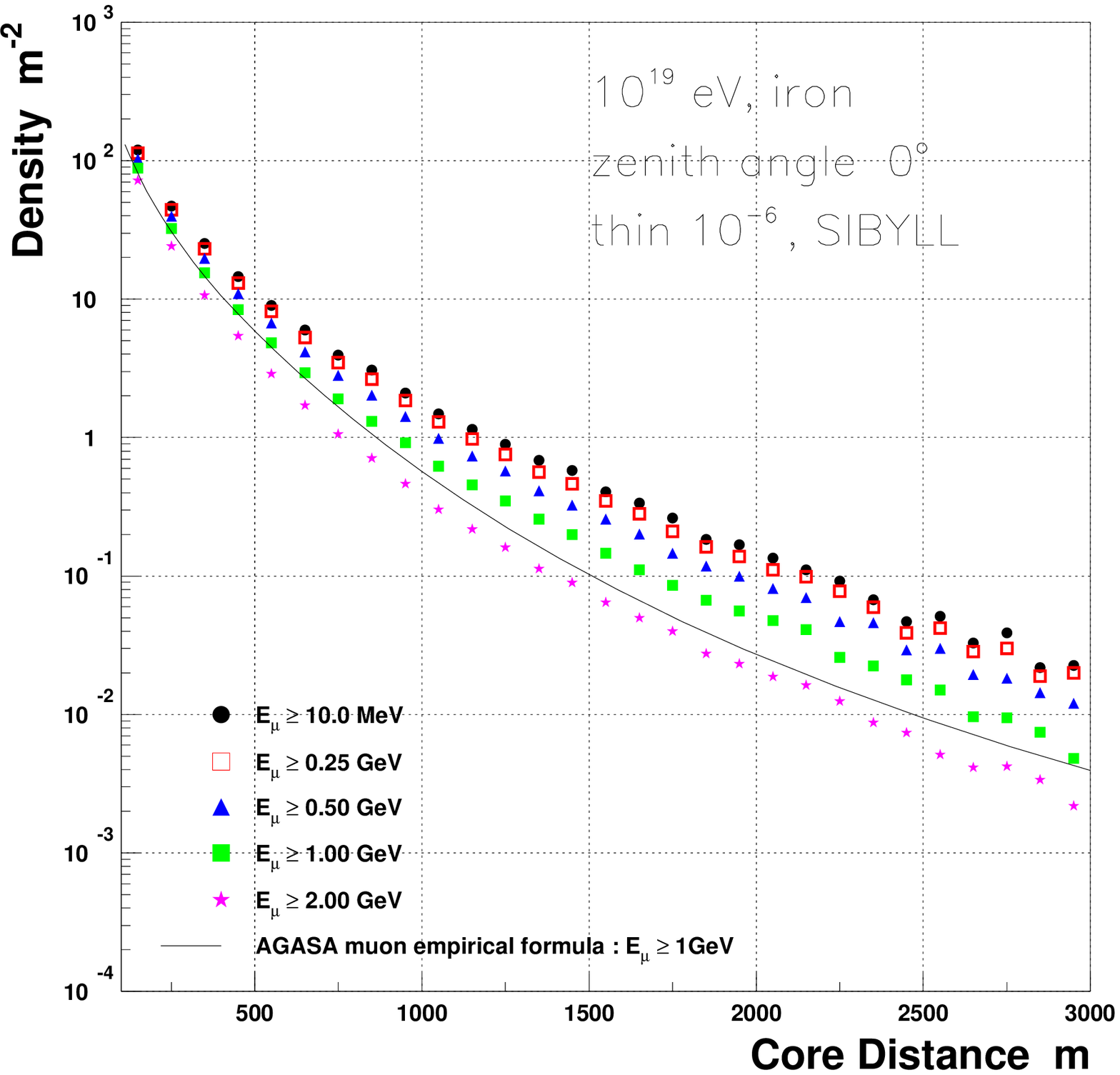,width=7cm,bbllx=0,bburx=522,bblly=18,bbury=520,clip=}
\end{center}
\caption{\small Same as Fig. \ref{mulat1}, except for the hadronic interaction model
is being SIBYLL.}
\label{mulat2}
\end{figure}

In Figs. \ref{mulat1} and \ref{mulat2}, lateral distributions of muons
above 10 MeV, 0.25 GeV, 0.5 GeV, 1 GeV and 2 GeV are compared with the
experimental formula of threshold energy of 1 GeV (Eq. \ref{eq-muon}) at
10$^{18}$ eV (left) and 10$^{19}$ eV (right).  The lateral distributions
become steeper as the cutoff energy increases.  While for QGSJET
calculations proton induced showers describe the experimental curve
best, for SIBYLL simulations iron induced showers are closer to the
experimental distributions.

It should be noted that the smaller muon number from the SIBYLL model
relative to other interaction models used in CORSIKA has been pointed
out in the 10$^{14}$~eV $\sim$ 10$^{15}$ eV region by Knapp {\it et al.}
\cite{knapp}.  The slope of energy spectra of muons between 0.25 GeV and
1.5 GeV at 600 m from the core is similar irrespective of the QGSJET and
the SIBYLL models and proton and iron primaries for 10$^{19}$ eV.  The
experimental result \cite{matsu85} agrees well with the simulated ones
except at 0.25 GeV, where we have in the experiment punch-through of the
electromagnetic component into the muon detectors, even at 600 m from
the core.

The relation of muon density above 1 GeV at 600 m from the core
($\rho_{\mu}(600)$) with the primary energy for three different
combinations of simulation codes and interaction models is compared in
Table \ref{tab-mu-e}.  Here the slope $\alpha$ in $\rho_{\mu} \propto
E^{\alpha}$ and the density at 10$^{19}$~eV are listed.  The results
from the SIBYLL model with the MOCCA code \cite{dawson98} are also
listed for comparison.  Though the hadronic model is the same, there is
about 10\% difference in density between the results based on the two
different simulation codes.  The implications from this comparison will
be made in Section 5.4.
\begin{table}
\caption{\small Comparison of slope and density at 10$^{19}$ eV}
\medskip
\begin{center}
\begin{tabular}{llllcl} \hline
Code & Model & Primary & Slope & $\rho_{\mu}$(600) & Note \\ 
  &  &  &  & at 10$^{19}$ eV &  \\ 
\hline 
CORSIKA & QGSJET & proton & 0.92  &  3.85 & \\  
CORSIKA & QGSJET & iron   & 0.89  &  5.64 & \\  
CORSIKA & SIBYLL & proton & 0.88  &  2.39 & \\  
CORSIKA & SIBYLL & iron   & 0.87  &  3.96 & \\  
MOCCA   & SIBYLL & proton & 0.90  &  2.95 & [17] \\  
MOCCA   & SIBYLL & iron   & 0.86  &  4.57 & [17] \\  
\hline
\end{tabular}
\end{center}
\label{tab-mu-e}
\end{table}

\subsection{Charged particle density, $S_0$(600)}

The relation of $S_0$(600) with the primary energy is examined for
various conditions.  $S_0$(600) does not depend on interaction model
(QGSJET, SIBYLL) nor primary mass (proton, iron) within 20\% which
supports the previous simulations by Hillas {\it et al.} \cite{hillas} and Dai
{\it et al}. \cite{dai88}.  The $S_0$(600) vs. primary energy relation
is almost linear, irrespective of primary mass (proton or iron) in case
of QGSJET. However, $S_0$(600) $\sim E^{0.97}$ in case of SIBYLL.  There
are some differences between the simulated number of charged particles
and the relation used at AGASA.  The details will be discussed in
Section 5.3.

\subsection{Zenith angle dependence of charged particles}
\begin{figure}
\begin{center}
\epsfig{file=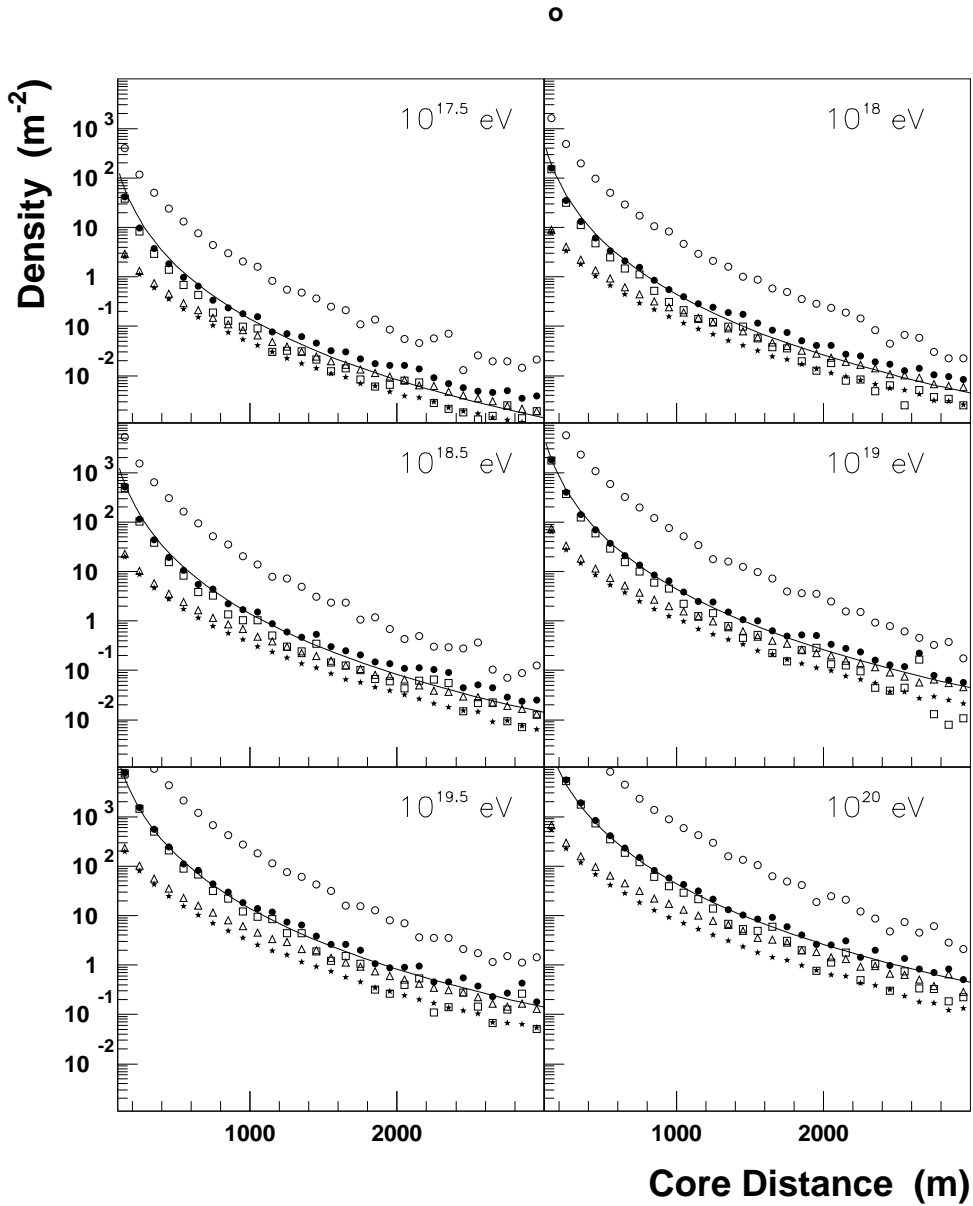,width=13cm,bbllx=7,bburx=423,bblly=14,bbury=500,clip=}
\end{center}
\caption{\small The lateral distributions of photons, electrons, muons and
charged particles simulated at a zenith angle of 39.7$^{\circ}$.  The
thinning level is 10$^{-6}$ and the cutoff energy of muons is 10 MeV and
those of electrons and photons are 1 MeV. For drawing the AGASA lateral
distribution, an attenuation length of 500 g cm$^{-2}$ is used and the
change of $\eta$ with zenith angle is taken into account.  Symbols as in
Fig. \ref{phelmu-6}.}
\label{fit-6-p-q-40}   
\end{figure}

There is a zenith angle dependence of the lateral distribution of
charged particles in the AGASA experiment and the slope parameter $\eta$
is expressed by Eq. \ref{eq-eta} as a function of zenith angle.  In
Fig. \ref{fit-6-p-q-40}, the lateral distribution of charged particles
at a zenith angle of 39.7$^{\circ}$ is shown for proton primaries of six
primary energies.  For the AGASA lateral distribution, $\eta = 3.6$ and an
attenuation length $\Lambda_{att}=500$ g cm$^{-2}$ are used.  It is
shown that the simulated results fit well to the experimental function,
including the absolute values, independent of primary energy.

\begin{figure}
\begin{center}
\epsfig{file=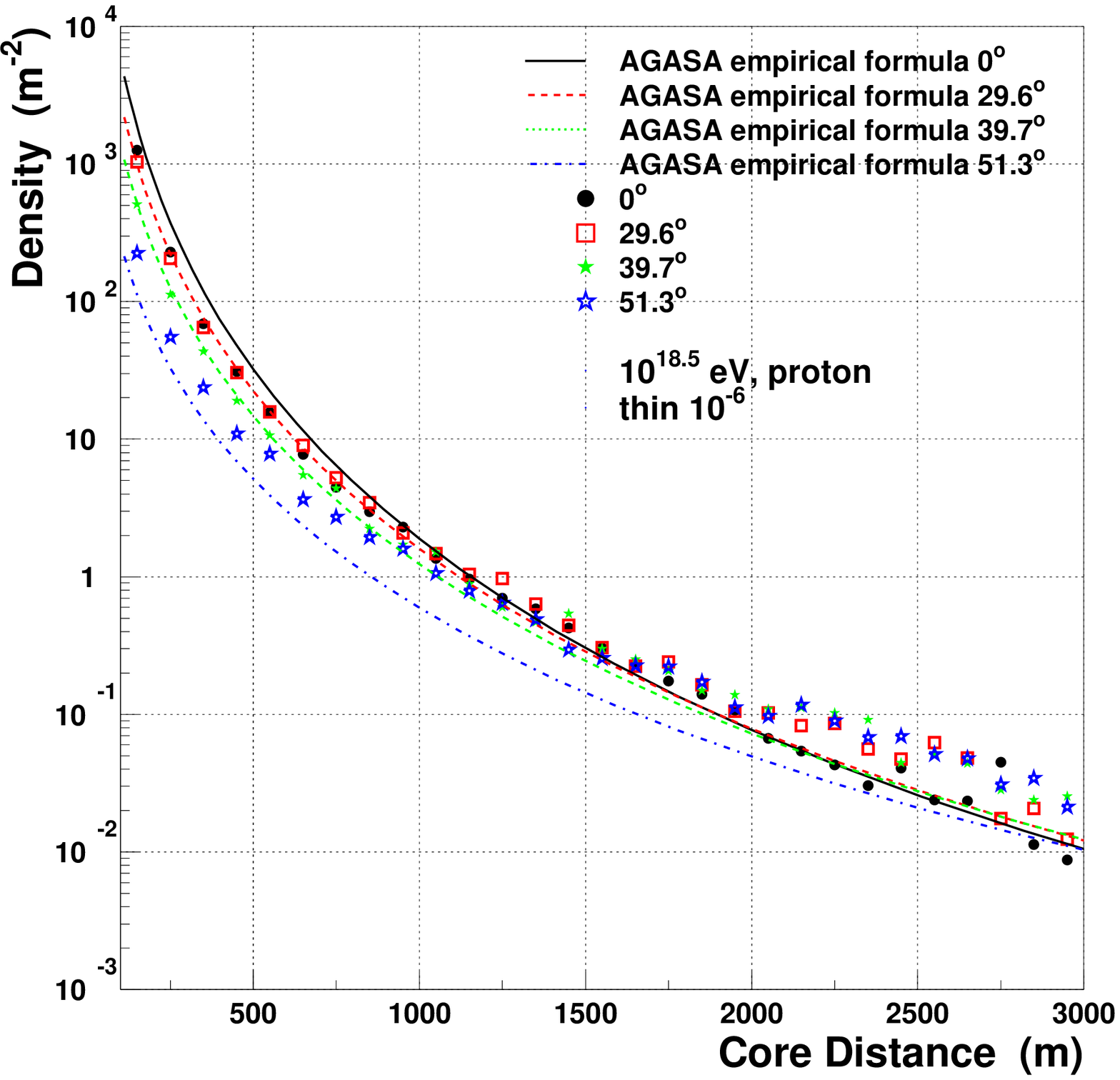,width=7cm,bbllx=0,bburx=526,bblly=14,bbury=522,clip=}
\hfill
\epsfig{file=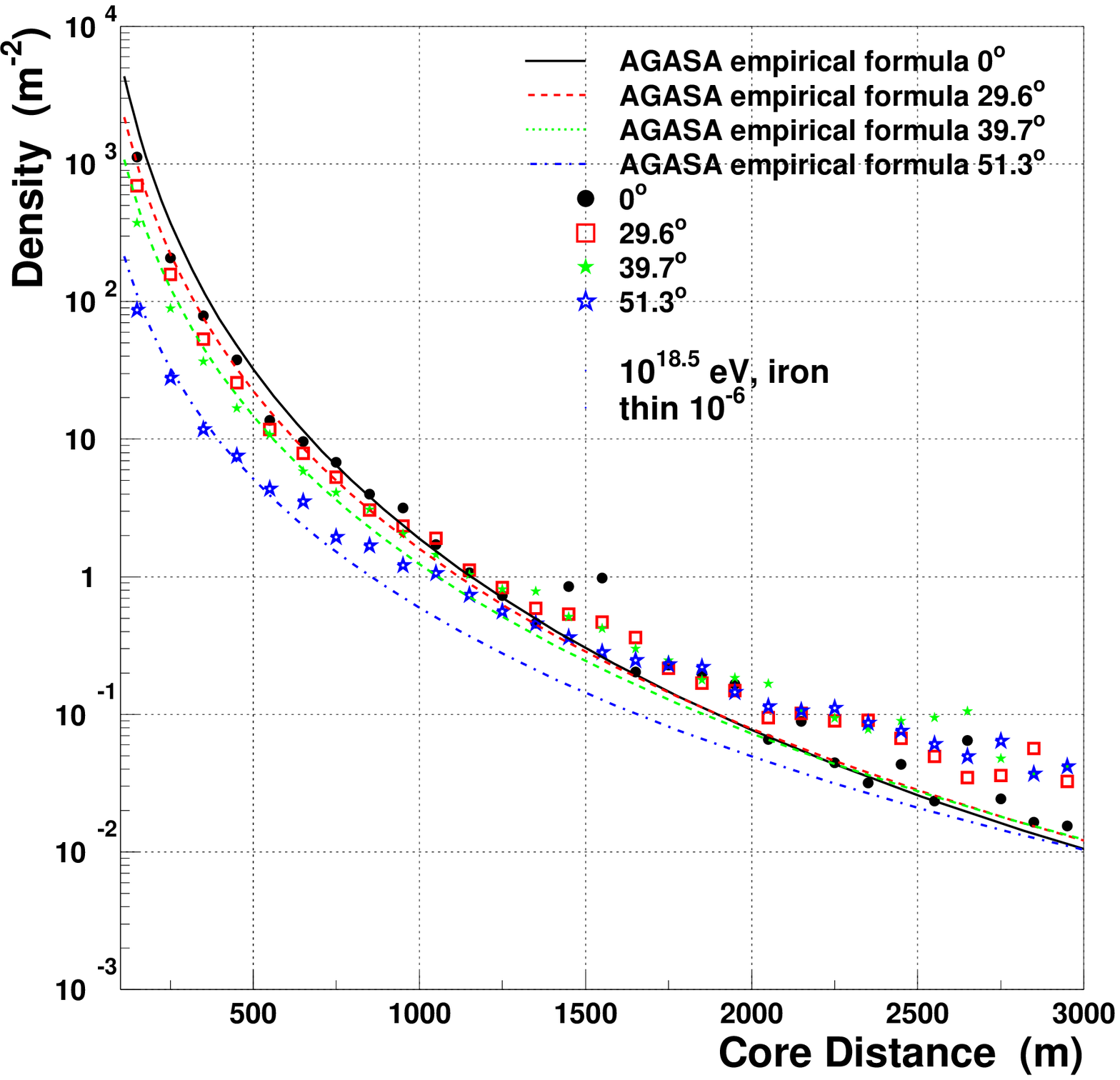,width=7cm,bbllx=0,bburx=526,bblly=14,bbury=522,clip=}
\epsfig{file=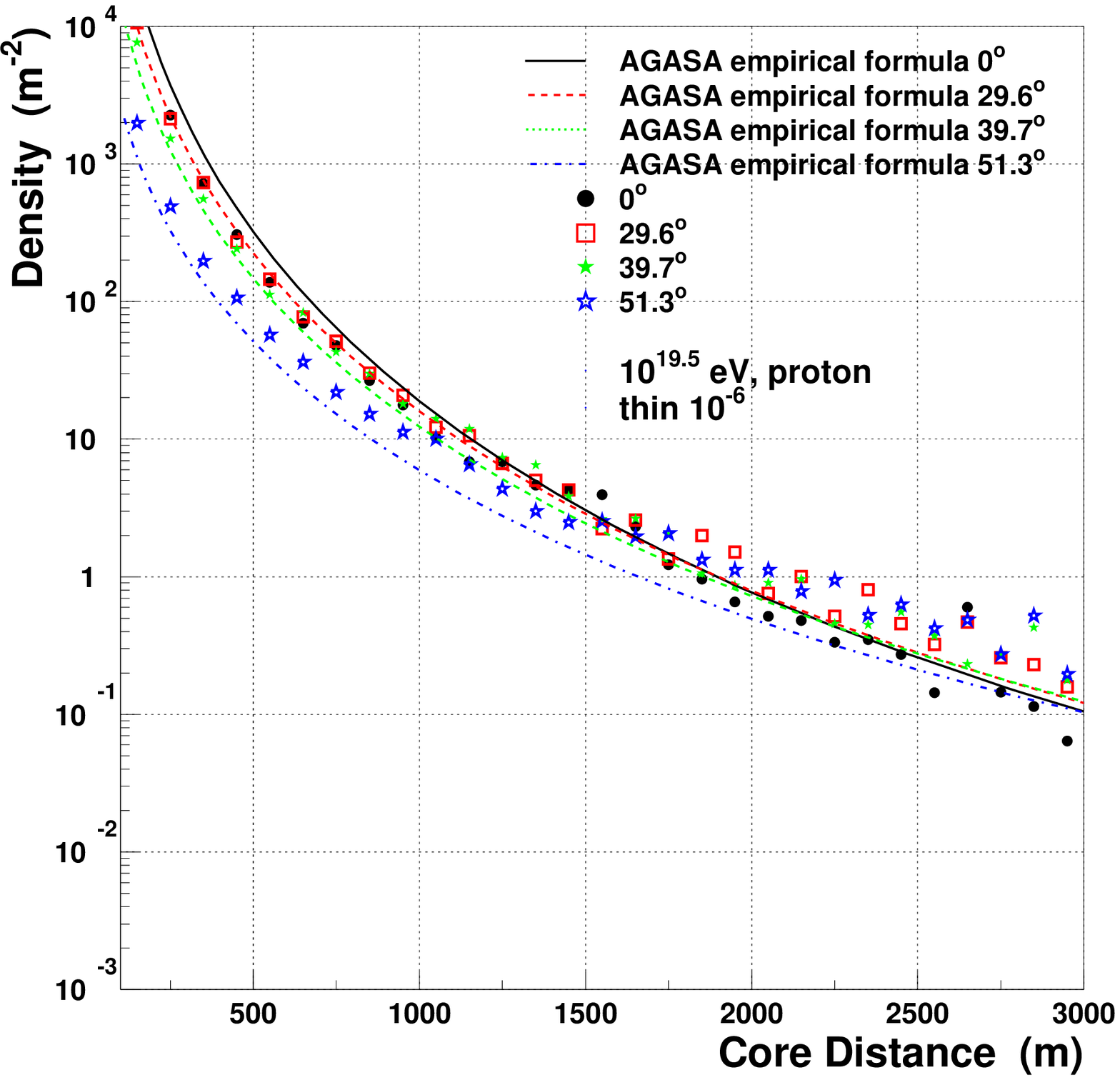,width=7cm,bbllx=0,bburx=526,bblly=14,bbury=522,clip=}
\hfill
\epsfig{file=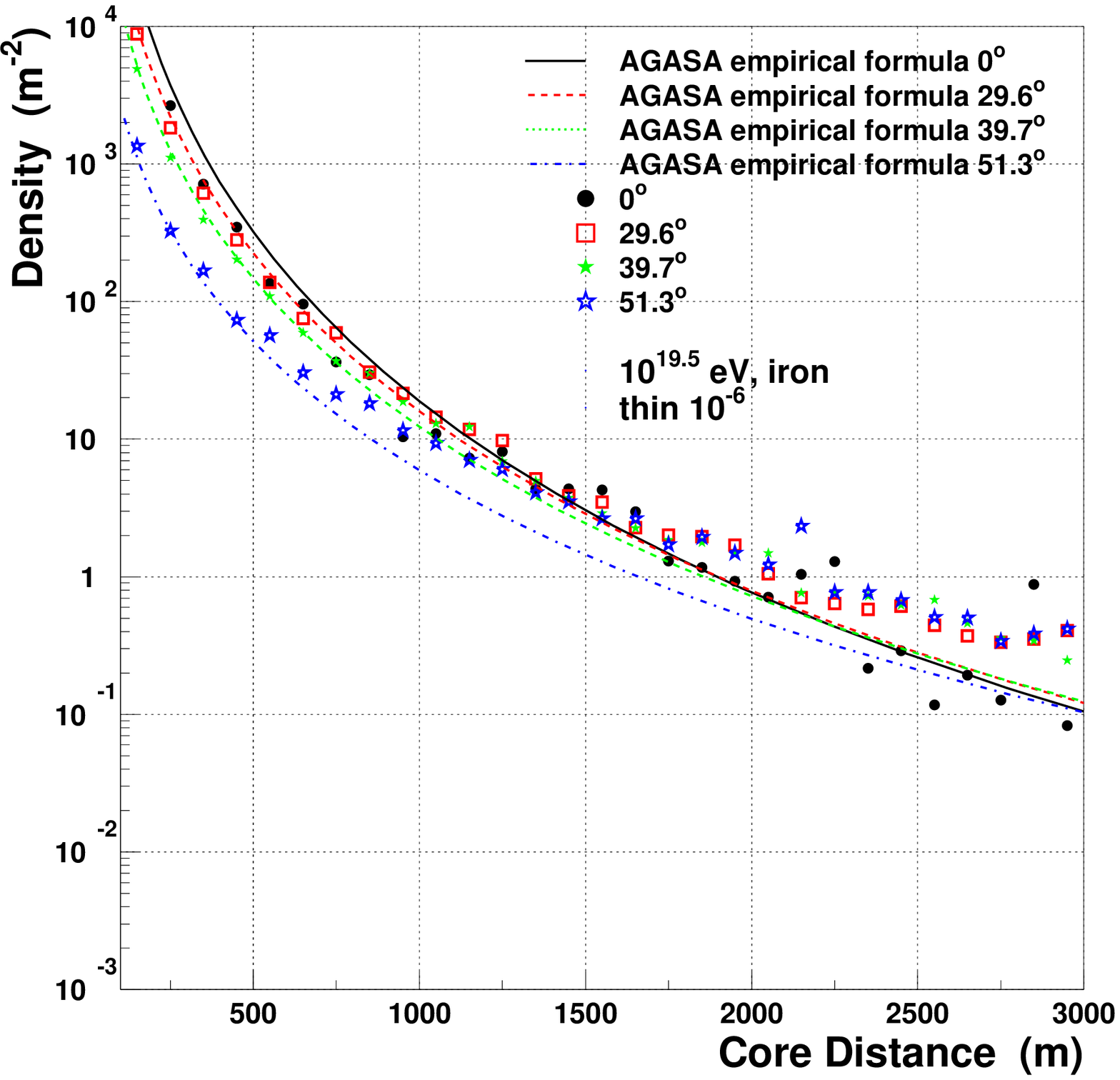,width=7cm,bbllx=0,bburx=526,bblly=14,bbury=522,clip=}
\end{center}
\caption{\small The lateral distributions of charged particles at four
zenith angles 0$^{\circ}$, 29.6$^{\circ}$, 39.7$^{\circ}$, and
51.3$^{\circ}$ for primary energies 10$^{18.5}$ eV (top) and 10$^{19.5}$
eV (bottom).  QGSJET model and cutoff energy of muons is 10 MeV and
those of photons and electrons are 1.0 MeV.  The left figure is for
proton primary and the right one is for iron primary.}
\label{att1}
\end{figure}

Figure \ref{att1} shows the lateral distributions of charged particles
at the four zenith angles $0^{\circ}$, 29.6$^{\circ}$, 39.7$^{\circ}$
and 51.3$^{\circ}$ for proton (left) and for iron (right) primaries and
for primary energies 10$^{18.5}$ eV and 10$^{19.5}$ eV, respectively.
It should be noted that there are no differences in attenuation of
charged particles with zenith angle between proton and iron primaries.
The details will be discussed in Section 5.5.

\section{Discussion}
The general features of the electromagnetic component and the low energy
muons observed by AGASA can be well reproduced by CORSIKA simulation,
except the slight differences in the absolute values.  These
discrepancies in absolute values between experiment and simulation are
partly due to the assignment of primary energy, which depends on the
definition of a single particle used in experiment and simulation.  More
remarkable discrepancies are the slope of $\rho_{\mu}$ vs. $S_0$(600)
relation and zenith angle dependence $S_{\theta}$(600) for constant
primary energy, which are shown in Fig. \ref{nmu-s600} and in
Fig. \ref{equi_sc600}, respectively.  In the following we evaluate the
simulated results with a similar definition of each observable as in the
AGASA experiment.

\subsection{Definition of density used in Akeno/AGASA experiment}
In the AGASA experiment, a scintillator of 5 cm thickness is used to
detect particles on the surface.  The scintillator is placed inside an
enclosure made of iron of 1.5 $\sim$ 2 mm thickness and the detector is
in a hut whose roof is made of an iron plate of 0.4 mm thickness.

The definition of a single particle at the Akeno experiment (V$_1$) is
the {\em average} of the pulse-height distribution of muons traversing a
scintillator vertically \cite{nagano84}. V$_1$ is 10\% larger than the
peak value, since the distribution is not Gaussian, but subject to
Landau fluctuations.  If we use the peak value in pulse {\em height}
distribution (PHD) of omnidirectional particles, the peak value V$_{ph}$
is accidentally coincident with V$_1$ at Akeno level (V$_{ph} \simeq$
V$_1$).  To convert a density measured by a scintillator to an electron
density corresponding to the calculated density by the NKG function
\cite{greisen56}, the density measured by scintillators and spark
chambers at the Institute for Nuclear Study at Tokyo \cite{shibata65} is
used at Akeno. This ratio is 1.1 between 10 m and 100 m from the core
and hence the electron shower size (N$_e$) determined by the Akeno 1
km$^2$ array was reduced by 10\% from the calculated N$_e$
\cite{nagano84}.

In AGASA a peak value V$_{pw}$ in the pulse {\em width} distribution
(PWD) of omnidirectional muons on a 5 cm scintillator is used as a
single particle conventionally.  The pulse width is obtained by
discriminating a signal, which decays exponentially with a time constant
of 10 $\mu$sec, at a constant level.  V$_{pw}$ is not equal to V$_{ph}$
and is related to V$_{ph}$ as :
\begin{displaymath}
  V_{pw}=(V_{ph} + \sqrt{V_{ph}^2 + \sigma^2})/2 
\end{displaymath}
where $\sigma$ is a full width at half maximum in PHD \cite{akeno}. By
putting V$_{ph}=1.0$ and $\sigma =0.70$, we obtain V$_{pw}=1.1$.  A
conventional value, V$_{pw}$, used at AGASA corresponds to
1.1$\times$V$_1 (\simeq 1.1\times$V$_{ph}$).  That is V$_{pw}$
corresponds to a measured density by spark chamber or the electron
density, as far as the ratio of the density measured by scintillators
and the spark chambers is 1.1.

Though there is a transition effect of the electromagnetic component in
scintillator of 5 cm thickness within 30 m from the core
\cite{nagano84}, the density of charged particles as expressed in units
of V$_1$ does not depend on the thickness of the scintillator above 30 m
from the core as shown experimentally in Teshima {\it et al}.
\cite{teshima86}.  This can be understood since the radiation lengths of
scintillator and air are very similar, so that the fraction of electrons
at any depth in the scintillator changes only slowly as compared to
air. This independence of thickness of scintillator has also been shown
in the simulations of Cronin \cite{cronin97} and Kutter \cite{kutter98}.

Assuming the 10\% difference of scintillator density to spark chamber
density at 600 m from the core, the density in units of V$_{pw}$ (1.1
$\times$ V$_{ph}$) coincide with spark chamber density as described
above and Eq. \ref{eq-energy1} was applied to deriving the AGASA energy
spectrum.

\subsection{Densities measured by scintillator of
5cm thickness}

So far the AGASA group has used a factor 1.1 of scintillator density to
spark chamber density which is determined within 100m from the core
\cite{shibata65}. However, the factor has not yet been measured beyond
100 m from the core.  In the following we discuss the lateral
distribution of energy losses by photons, electrons and muons in 5cm
scintillator in units of V$_{pw}$ at Akeno level, taking into account
the incident angles of electrons and photons far from the core, because
the incident angles of electrons and photons on scintillator may not be
vertical even for a vertical shower and the particles have some angular
distribution.

\begin{figure}
\begin{center}
\epsfig{file=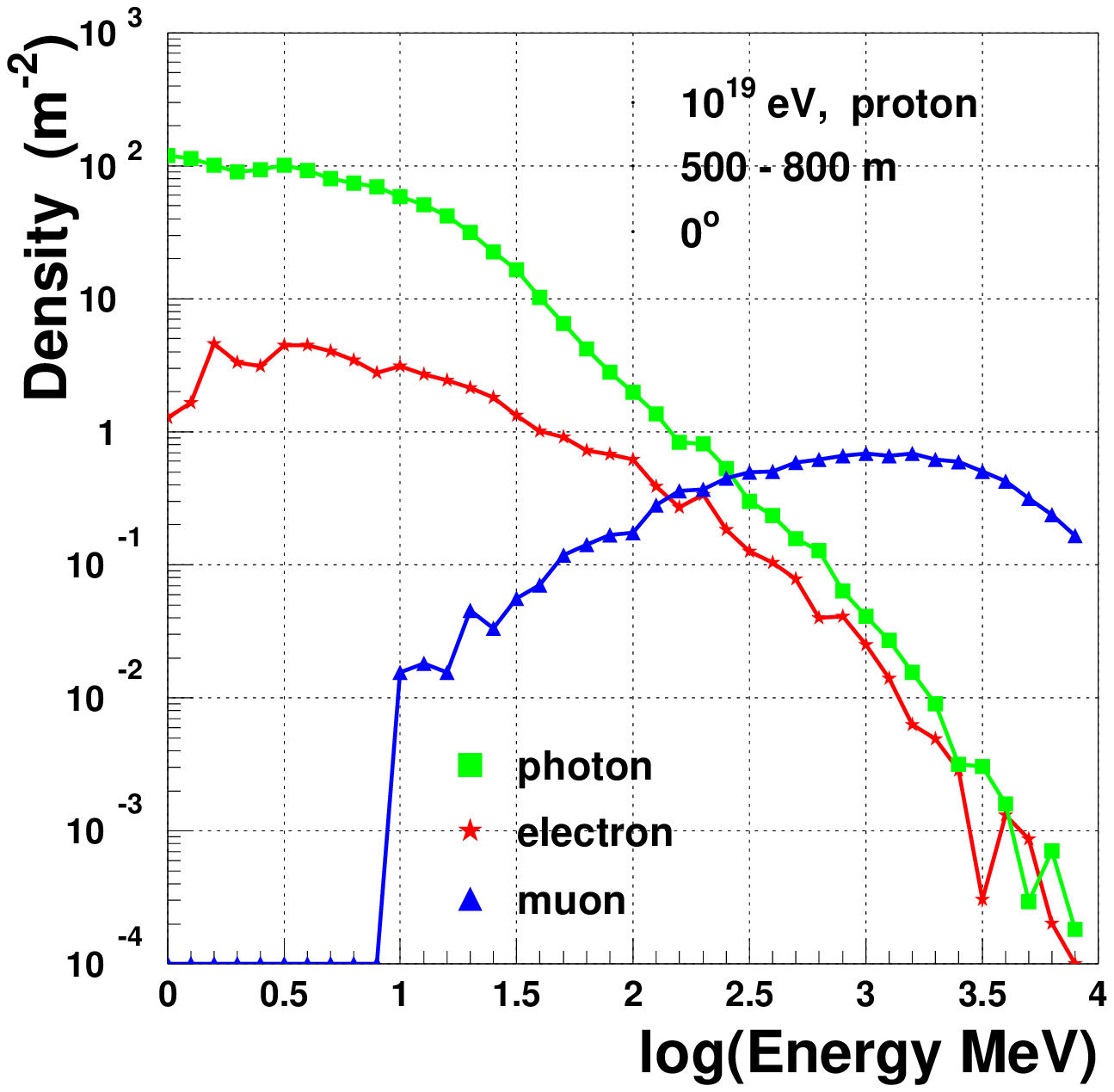,width=7cm,bbllx=1,bburx=373,bblly=14,bbury=378,clip=}
\hfill
\epsfig{file=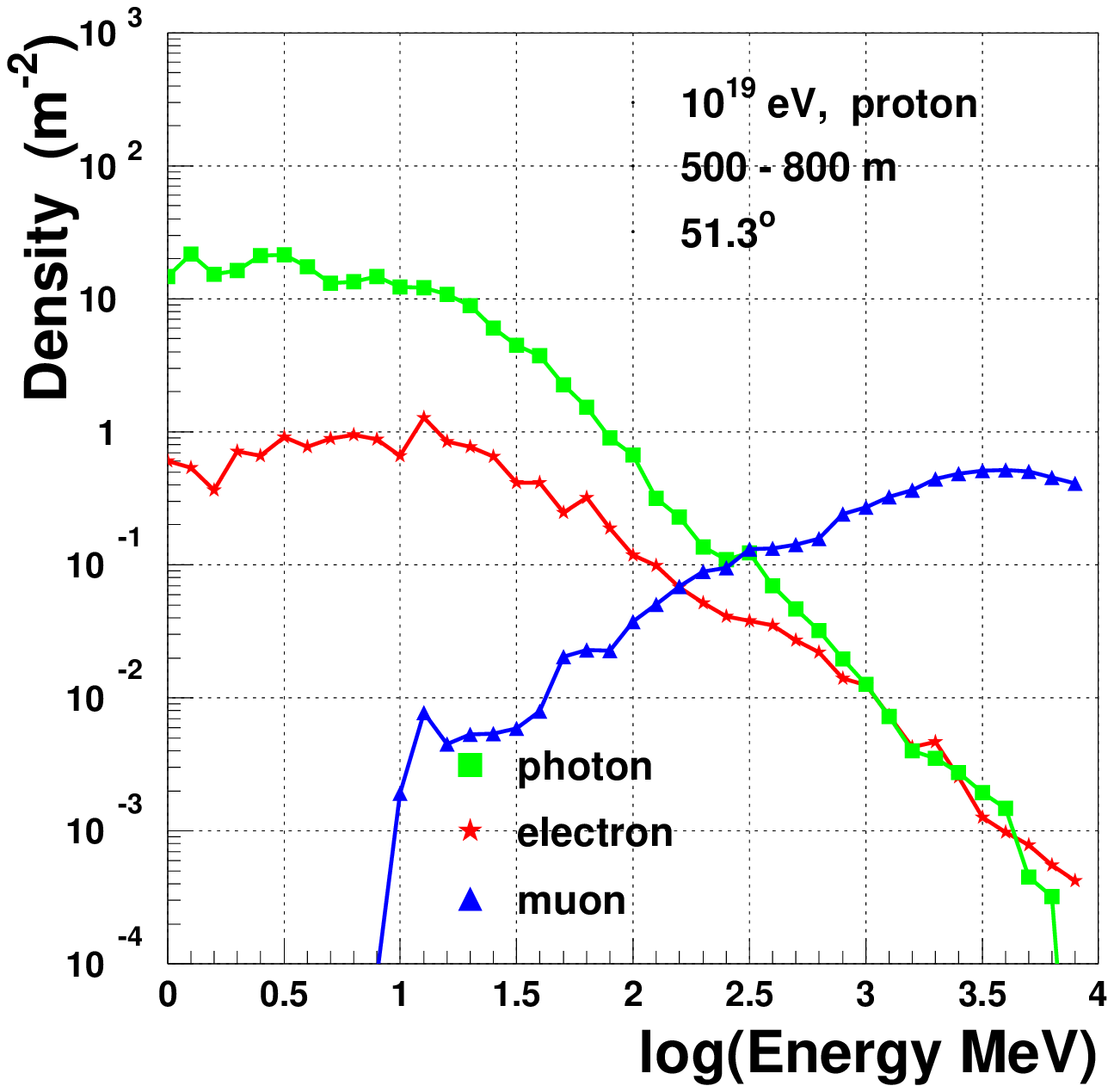,width=7cm,bbllx=1,bburx=373,bblly=14,bbury=378,clip=}
\end{center}
\caption{\small Energy spectra of photons, electrons and muons at core
distance between 500 m and 800 m.  The primary particle is a proton of
10$^{19}$ eV. Zenith angles are 0$^{\circ}$ (left) and 51$^{\circ}$
(right).}
\label{egy600}
\end{figure}

In Fig. \ref{egy600} the energy spectra of photons, electrons and muons
are shown at core distances between 500 m and 800 m, simulated by
CORSIKA.  Many photons still remain at a zenith angle of 51.3$^{\circ}$
($\sec\theta =1.6$).

Using CORSIKA, in a shower the incident angle of each particle to the
surface is recorded, so that we can calculate the energy loss of each
particle which is incident on the scintillator with various zenith
angles.  In the case of electrons and muons, only ionization loss is
taken into account.  The energy loss of photons is evaluated as follows.
By dividing the scintillator in thin layers, the fraction of conversion
to electrons in each layer is calculated using the photon attenuation
length in water \cite{barnett96}.  The energy loss of electron or
electron/positron in the remaining layers for the fraction of photons is
calculated.  In this way the average energy loss of a single photon in a
scintillator of various thicknesses is calculated as a function of the
photon energy.  Though the calculation process is simple, the result for
a scintillator of 5 cm thickness agrees well with the Monte Carlo
simulation results by Kutter \cite{kutter98}.  The density of a shower
of zenith angle $\theta$ is evaluated as the energy loss in scintillator
of $5\times\sec\theta$ cm thickness (in units of V$_{pw}$) in an area of
$1\times\cos\theta$ m$^2$.

\begin{figure}
\begin{center}
\epsfig{file=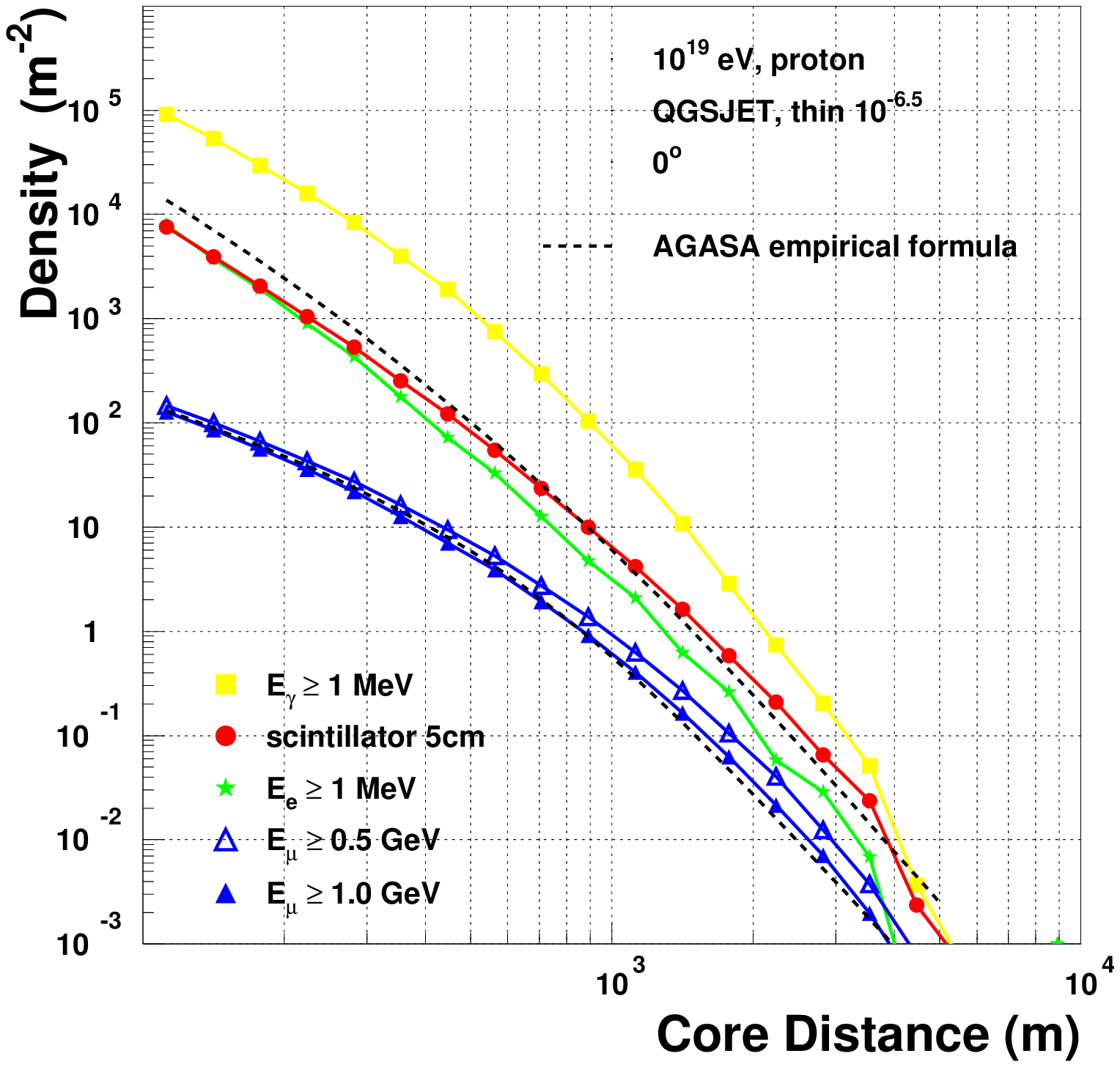,width=7cm,bbllx=1,bburx=440,bblly=10,bbury=428,clip=}
\hfill
\epsfig{file=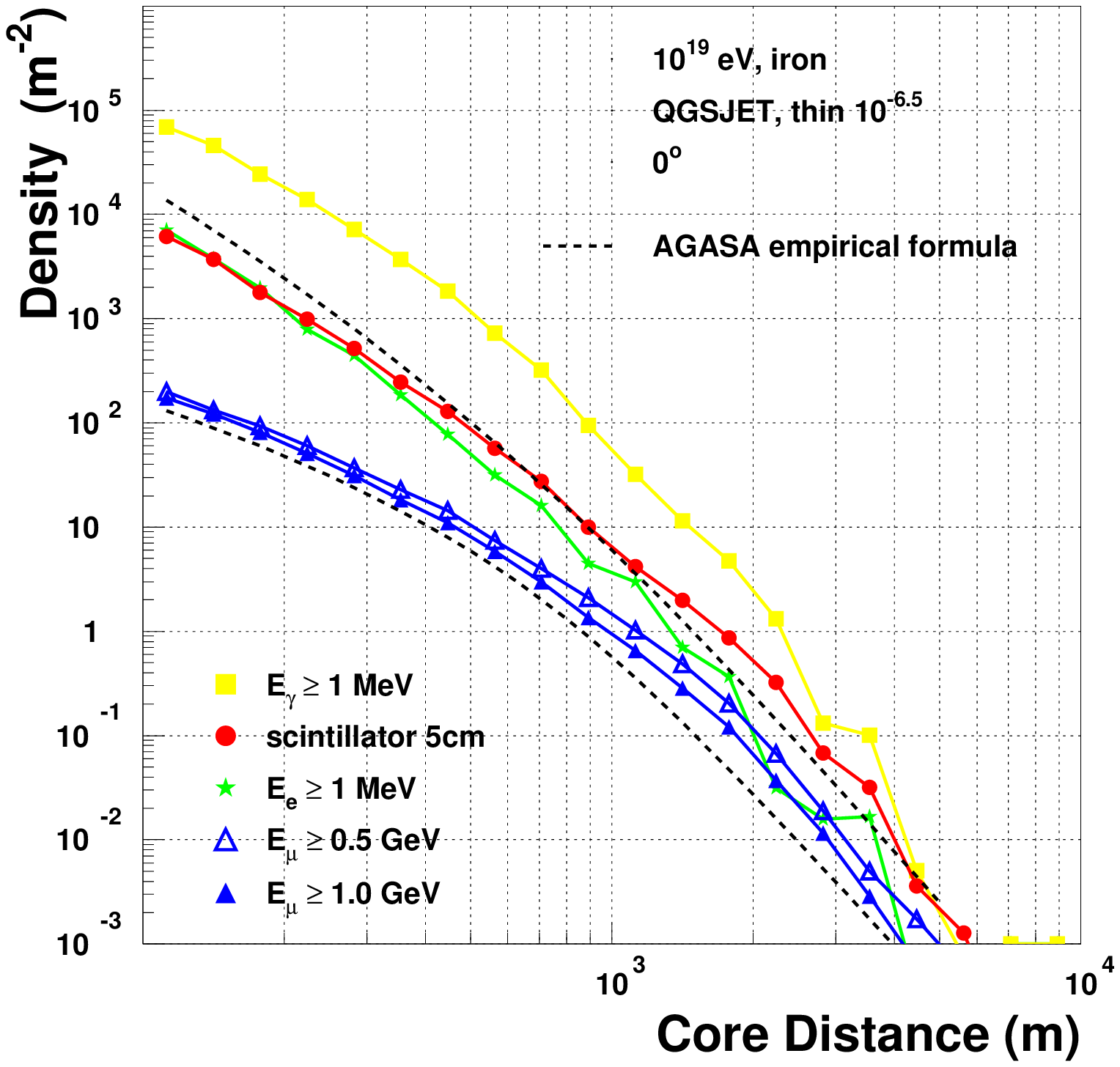,width=7cm,bbllx=1,bburx=440,bblly=10,bbury=428,clip=}
\end{center}
\caption{\small  Lateral distribution of energy deposit in scintillator
($\bullet$) of 5 cm thickness ($\rho_{sc}$) in units of V$_{pw}$ is
compared with the experimental lateral distribution of AGASA.  Those of
electrons ($\ge$ 1 MeV), photons ($\ge$ 1 MeV), muons ($\ge$ 0.5 GeV)
and muons ($\ge$ 1 GeV) are also shown.  The left figure is proton
primary and the right one iron.}
\label{lat_egy0}
\end{figure}

In Fig. \ref{lat_egy0}, the lateral distribution of energy deposit in
scintillator of 5 cm thickness in units of V$_{pw}$ ($\rho_{sc}$) is
plotted by closed circles and compared with that of the experimental
lateral distribution (dashed line).  The simulated lateral distribution
is flatter than the experimental one.  $\rho_{sc}$ reflects the number
of electrons near the core (up to about 200 m from the core), but
becomes larger than the electron density with core distance.\ Though the
agreement of $\rho_{sc}$(600) with the experimental $S_0$(600) is quite
good, the difference of the lateral distribution of $\rho_{sc}$ from the
experiment must be studied further.

\subsection{Primary energy and $S_0$(600) relation evaluated by CORSIKA}
In Table \ref{s600}, the density of charged particles at 600 m from the
core, $\rho_{ch}$(600), or the scintillator density of 5 cm thickness,
$\rho_{sc}$(600), are listed for showers of 10$^{19}$ eV of vertical
incidence.
\begin{table}[thb]
\caption{\small Comparison of charged particle density ($\rho_{ch}$)
or scintillator density in units of V$_1$ ($\rho_{sc}$)
at 600 m from the core 
for showers of vertical incidence, 10$^{19}$ eV and proton or iron primary.}  
\medskip
\begin{tabular} {lllcccl} \hline
Code & Model  & Primary & Thinning & Threshold &
 $\rho_{ch}$(600) or & Note\\ 
 &  &  & level  & E$_{e\gamma}$ (MeV)& $\rho_{sc}$(600) (m$^{-2}$) & \\ 
\hline
\hline
CORSIKA & QGSJET & proton &  10$^{-5}$  &  1.0   &  32.5 & \\ \hline
CORSIKA & QGSJET & proton &  10$^{-6}$  &  1.0   &  32.4 & \\ \hline
CORSIKA & QGSJET & proton &  10$^{-6}$  & 0.1   & 37.5 & \\ \hline
CORSIKA & QGSJET & proton &  10$^{-6}$  & 0.05   & 39.1 & \\ \hline
CORSIKA & QGSJET & iron   &  10$^{-5}$  &  1.0   &  35.7  & \\ \hline
CORSIKA & QGSJET & iron   &  10$^{-5}$  & 0.05   & 39.8  & \\ \hline
CORSIKA & SIBYLL & proton &  10$^{-6}$  &  1.0   &  30.4 &  \\ \hline 
CORSIKA & SIBYLL & iron   &  10$^{-6}$  &  1.0   & 33.5  &  \\ \hline
CORSIKA & QGSJET & proton &  10$^{-6}$  &  1.0   & 38.0  &  \\ 
        & -NKG    &        &             &  0   & 45.0  &  \\ \hline 
CORSIKA & QGSJET & proton & 10$^{-6}$ & 1.0  & sci. 43.0 & \\ \hline 
CORSIKA & QGSJET & iron & 10$^{-5}$ & 1.0  & sci. 46.2  & \\ \hline 
CORSIKA & SIBYLL & proton & 10$^{-6}$ & 1.0  & sci. 38.2  & \\ \hline 
CORSIKA & SIBYLL & iron & 10$^{-6}$ & 1.0  & sci. 44.4  & \\ \hline \hline 
MOCCA & SIBYLL & proton &     & 0.1   &  33.5 & Cronin $^{(1)}$ \\ 
  & & iron &     &  0.1   &  38.7 & Cronin $^{(1)}$ \\ \hline \hline
COSMOS & QCDjet & proton & 10$^{-5}$ & 0  & 50.0  & Dai et al.$^{(2)}$ \\ 
 & -NKG  & CNO &  10$^{-5}$  & 0  &    &  \\ 
 &  & iron &  10$^{-5}$  & 0   &    &  \\ \hline 
\end{tabular}
\noindent{$^{(1)}$ \footnotesize Simulation results made at Fermi
Lab. using the SIBYLL interaction model with MOCCA simulation code
(J.Cronin \cite{cronin97})}.\\
\noindent{$^{(2)}$ \footnotesize Two dimensional simulation results made
at ICRR with COSMOS by Dai {\it et al}. \cite{dai88}.  Photons and
electrons of energies below the thinning energy level are connected to
the NKG function in which the Moli\`{e}re length is used at 2 radiation
lengths above the Akeno altitude.}
\label{s600}
\end{table}

The various combinations of simulation codes (CORSIKA, MOCCA), hadronic
interaction models (QGSJET, SIBYLL), primary species (proton, iron),
thinning levels and threshold energies of the electromagnetic components
are compared.

In general, the difference of $\rho_{ch}$(600) due to the difference of
simulation codes or hadronic interaction models is within 10\% for the
same cutoff energy of electromagnetic component.  $\rho_{ch}$(600)
depends on the cutoff energy of electromagnetic component.  In the
$S_0$(600) vs. energy relation by Dai {\it et al.}
(Eq. \ref{eq-energy1}), the electromagnetic component with energy of
less than the thinning level is connected to the NKG function without
cutoff energy for electrons and photons, and MU at 2 radiation lengths
above the observation level is used.  The result by CORSIKA simulated
with the similar method is given in Table \ref{s600} as QGSJET-NKG.  In case
of CORSIKA, the relation is
\begin{equation}
  E [\mbox{eV}] = 2.2 \times 10^{17} S_0(600)^{1.0}
\label{eq_cor_energy}
\end{equation} 
and is about 10\% larger than that by Dai {\it et al}.

\begin{figure}
\begin{center}
\epsfig{file=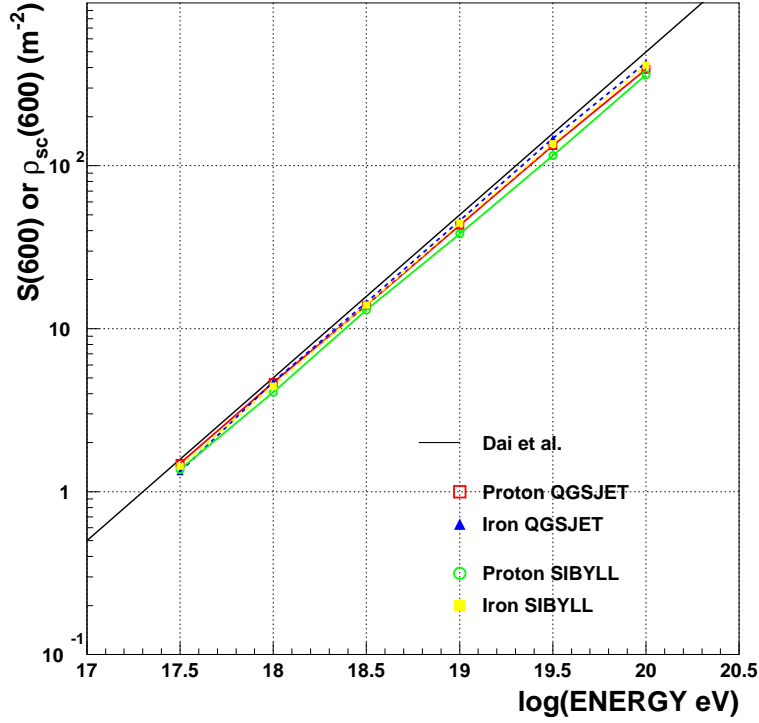,width=10cm,bbllx=1,bburx=524,bblly=11,bbury=513,clip=}
\end{center}
\caption{\small $S_0$(600) or $\rho_{sc}$(600) vs. energy relation
estimated from the QGSJET and SIBYLL model for proton and iron primary.}
\label{sc600_egy}
\end{figure}

The energy losses of photons, electrons and muons in scintillator of 5
cm thickness ($\rho_{sc}$(600)) in units of V$_{pw}$ have been evaluated
as described in the previous section by taking into account their
incident angle to the scintillator and attenuation of low energy photons
and electrons in the scintillator container and hut.  The relation is
drawn in Fig. \ref{sc600_egy} for proton and iron primaries. For the
QGSJET hadronic interaction model we obtain
\begin{eqnarray} 
\label{eq-qg-p}
 E [\mbox{eV}] &=& 2.07 \times 10^{17} \rho_{sc}(600)^{1.03} \quad \mbox{for proton}\\
 E [\mbox{eV}] &=& 2.24 \times 10^{17}  \rho_{sc}(600)^{1.00} \quad \mbox{for iron} 
\label{eq-qg-fe}
\end{eqnarray}  
and for SIBYLL the relation reads
\begin{eqnarray}   
\label{eq-si-p}
 E [\mbox{eV}] &=& 2.30 \times 10^{17}  \rho_{sc}(600)^{1.03} \quad \mbox{for proton}\\
 E [\mbox{eV}] &=& 2.19 \times 10^{17}  \rho_{sc}(600)^{1.01} \quad \mbox{for iron.} 
\label{eq-si-fe}
\end{eqnarray}   

Though there is a difference between proton and iron showers or QGSJET
and SIBYLL, any combination assigns a higher energy than that by
Eq. \ref{eq-energy1}.  The functions given by Eqs. \ref{eq-qg-p},
\ref{eq-qg-fe} and \ref{eq-si-fe} intersect around
$\rho_{sc}(600)=20\sim30$, the combined relation may be written as
\begin{eqnarray} 
 E [\mbox{eV}] &=& 5.65 \times 10^{18}
 \left(\rho_{sc}(600)\over{25.0} \right)^{1.015}, 
\label{eq-average}
\end{eqnarray}   
by taking the average slope of the four equations.  If we rewrite the
above equation in similar form as Eq. \ref{eq-energy1}, the following
function may be used.
\begin{equation} 
  E [\mbox{eV}]  = 2.15 \times 10^{17} S_0(600)^{1.015}.
\label{eq-corsika}
\end{equation} 
Here we denoted $S_0$(600) instead of $\rho_{sc}$(600) as a scintillator
density used in the experiment, since $S_0(600) \sim \rho_{sc}$(600) as
described before.  This equation means that the energy of the AGASA
events must be increased by 14\% at 10$^{19}$ eV and 18\% at 10$^{20}$
eV, if we evaluate the primary energy by CORSIKA.

\subsection{Composition deduced from the muon component}
\begin{figure}
\begin{center}
\epsfig{file=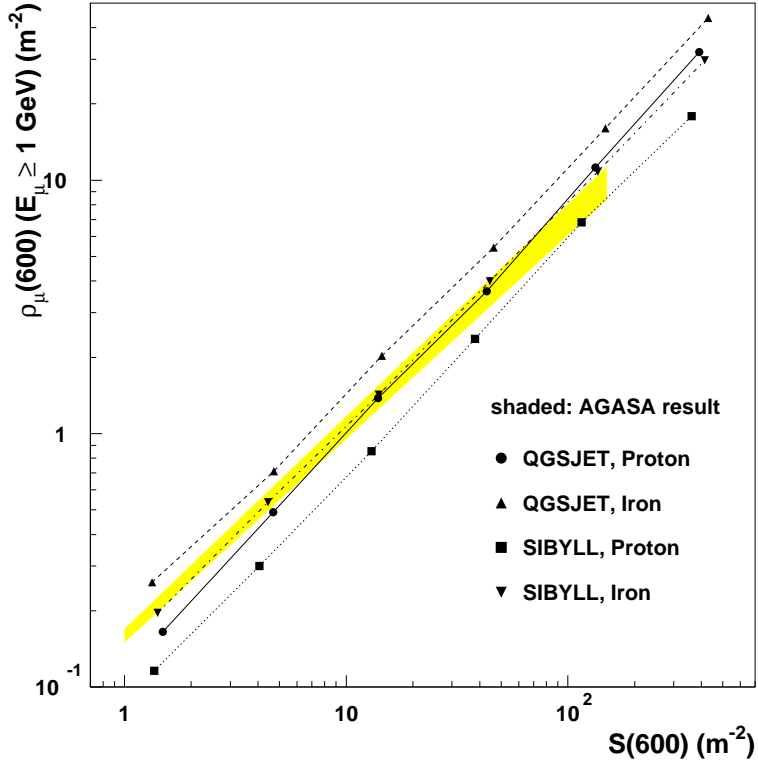,width=10cm,bbllx=9,bburx=554,bblly=4,bbury=555,clip=}
\end{center}
\caption{\small Comparison of experimental $\rho_{\mu}$ vs $S_0$(600) or
$\rho_{ch}$ relation with that from CORSIKA simulation.}
\label{nmu-s600}
\end{figure}

The slope of the $\rho_{\mu}$(600) vs. energy relation measured by
experiment is smaller than that in CORSIKA simulations.  In
Fig. \ref{nmu-s600} the $\rho_{\mu}$(600) vs. $S_0$(600) (or
$\rho_{sc}$(600)) relation is shown.  Since the relation $S_0$(600)
vs. energy is almost linear irrespective of primary composition or
interaction model used, the relation
$$  \rho_{\mu}(600) = A \times S_0(600)^b $$ is similar to 
$\rho_{\mu}$(600) vs. energy relation and 
the parameter b is similar to $\alpha$ in Table \ref{tab-mu-e}.

\begin{figure}
\begin{center}
\epsfig{file=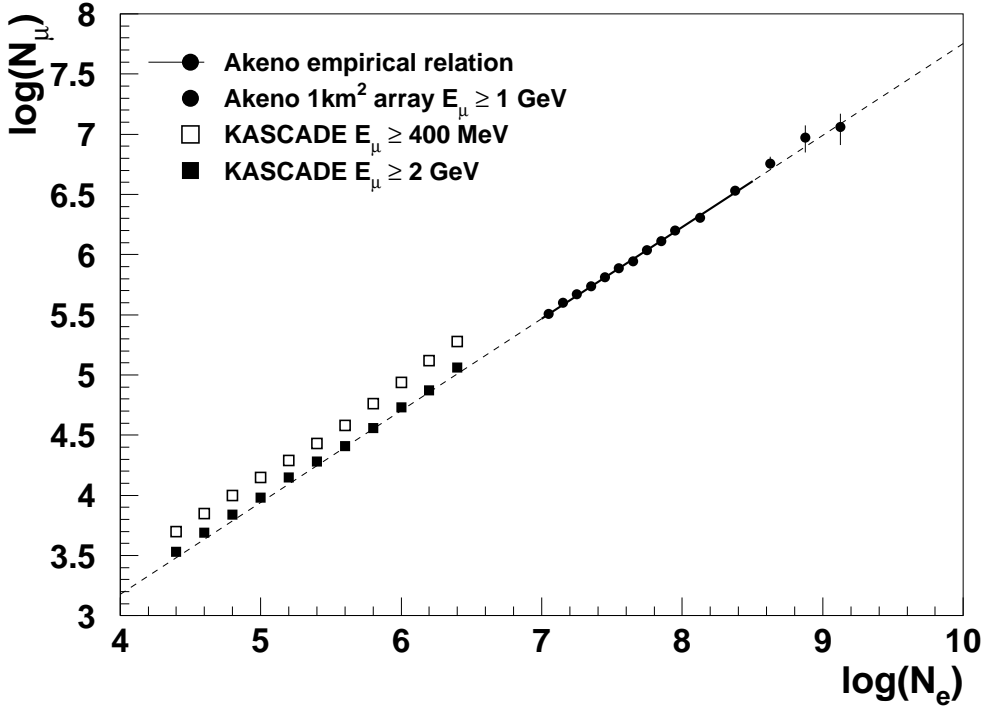,width=13cm,bbllx=1,bburx=461,bblly=13,bbury=347,clip=}
\end{center}
\caption{\small $N_{\mu}$ vs. $N_e$ relation of the Akeno and the KASCADE
experiments. The solid line represents Eq. \ref{eq-corsika} and the
dashed one is its extrapolation.}
\label{ne-nmu}
\end{figure}

To explain the difference of slopes between experiment and simulation,
Dawson {\it et al}. \cite{dawson98} favours a likely change in mass
composition from heavy to light at energies above 10$^{18}$ eV, as a
supporting evidence of the change of mass composition claimed by the
Fly's Eye experiment \cite{gaisser93}. The difficulty of their
interpretation arises when the relation is applied to the energy region
lower than 10$^{17}$ eV\footnote{The AGASA group claims that no change
of composition around 10$^{17.5}$ eV is observed by their experiment
\cite{hayashida95} in contrast to the Fly's Eye experiment. Dawson {\it
et al}. do not argue on this point (compare with Fig. 4 and 6 of their
paper).  According to their figure, the fraction of iron component is
100\% below 10$^{17.5}$ eV in the Fly's Eye experiment, while that is
80\% at 10$^{17.5}$ eV and exceeds 100\% below 10$^{17.0}$ eV in the
AGASA experiment. There is still no agreement between the two
experiments below 10$^{17.5}$ eV, which is the point in dispute.}.  The
$N_{\mu}$ vs $N_e$ relation of the Akeno result \cite{hayashida95} is
compared with the result of KASCADE \cite{leibrock98} in
Fig. \ref{ne-nmu} without any correction for each experiment.  The
empirical formula of Akeno at $\sec \theta =1.05$ is expressed by
\begin{equation}
 N_{\mu}=(2.94 \pm 0.14)\times10^5 \times (N_e/10^7)^{0.76\pm0.02}.
\label{eq-mu-ne}
\end{equation}
$N_e$ attenuates about a factor 1.5 from the Akeno level (920 g
cm$^{-2}$) to the KASCADE level (1020 g cm$^{-2}$), while $N_{\mu}$ of
the KASCADE experiment with the threshold energies of 2 GeV is expected
to be smaller than that at Akeno of 1 GeV about a factor $1.2\sim 1.5$
\cite{greisen60}.  Therefore such an agreement of the extrapolation of
the absolute values from both experiments is well understood.  The
important result, however, is the agreement of the slopes of the
$N_{\mu}$ vs. $N_e$ relation of both experiments in quite different
energy regions.

Since in the higher energy region, $N_e$ can not be determined by AGASA,
the $\rho_{\mu}$(600) vs. $S_0$(600) relation is used.  The result at
$\sec\theta=1.09$ \cite{hayashida95} is expressed as
\begin{equation}
 \rho_{\mu}(600)=(0.16\pm 0.01) \times S_0(600)^{0.82\pm0.03}.
\end{equation}
Taking into account the relation $S_0(600)\propto E_0 \propto
N_e^{0.9}$, the above equation coincides with Eq. \ref{eq-mu-ne} in the
overlapping energy region \cite{hayashida95}.  Therefore the slope seems
not to change from 10$^{14.5}$ eV to 10$^{19}$ eV.  If we take the
absolute values of SIBYLL model as used in Dawson {\it et al}.
\cite{dawson98} (refer to Fig. 4 of their paper), the composition
becomes heavier than iron below 10$^{17}$ eV.  This conclusion
demonstrates that it is important to compare the experimental results
with simulations in as wide energy range as possible.

\subsection{Attenuation of $S_0$(600) with zenith angle and the
implication on the primary composition around 10$^{19}$ eV}

\begin{figure}
\begin{center}
\epsfig{file=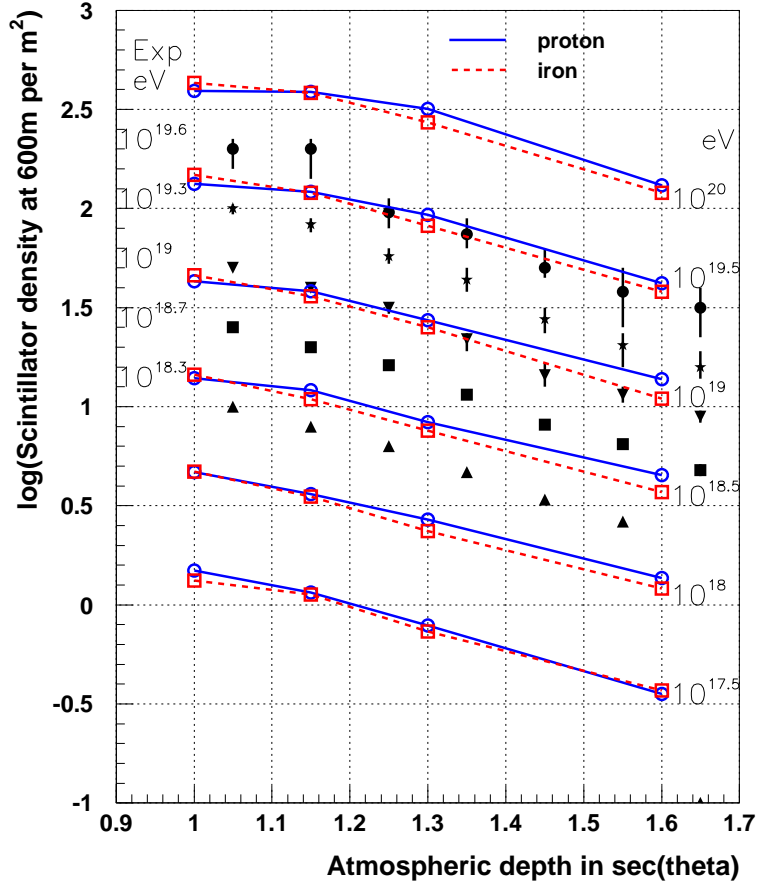,width=10cm,bbllx=1,bburx=377,bblly=16,bbury=460,clip=}
\end{center}
\caption{\small The variation of $\rho_{sc}$(600) with zenith angle.
Solid lines are from protons and dotted are from iron (QGSJET model,
thinning level 10$^{-6}$).  Experimental points [20] correspond to 40,
20, 10, 5 and 2 $\times10^{18}$ eV from top to bottom, respectively.}
\label{equi_sc600}
\end{figure}

The $\rho_{sc}$(600) values are plotted in Fig. \ref{equi_sc600} as a
function of atmospheric depth ($\sec\theta$) and connected with lines.
The solid lines represent proton primaries and dotted ones are iron
primaries.  The experimental points of AGASA \cite{hayashida97} are also
plotted.  These were obtained using the method of `equi-intensity cuts'
on the integral $\rho_{sc}$(600) spectra, based on the assumption that
the flux of showers above a certain primary energy does not change with
atmospheric depth.  As shown in the figure the variation of
$\rho_{sc}$(600) with zenith angle by CORSIKA (QGSJET) is similar to
proton and iron primaries and is smaller than that of the experiment.
If we take into account the error in $\rho_{ch}$(600) and zenith angle
determination, this difference becomes larger, since flux of the
observed $\rho_{ch}$(600) spectrum is increased by a factor of
$\exp(\sigma^2(\gamma-1)^2/2)$, where $\sigma$ is the error in
$\rho_{ch}$(600) determination in a logarithmic scale and $\gamma$ is
the power index of the differential $\rho_{ch}$(600) spectrum
\cite{murzin65}.

\begin{figure}
\begin{center}
\epsfig{file=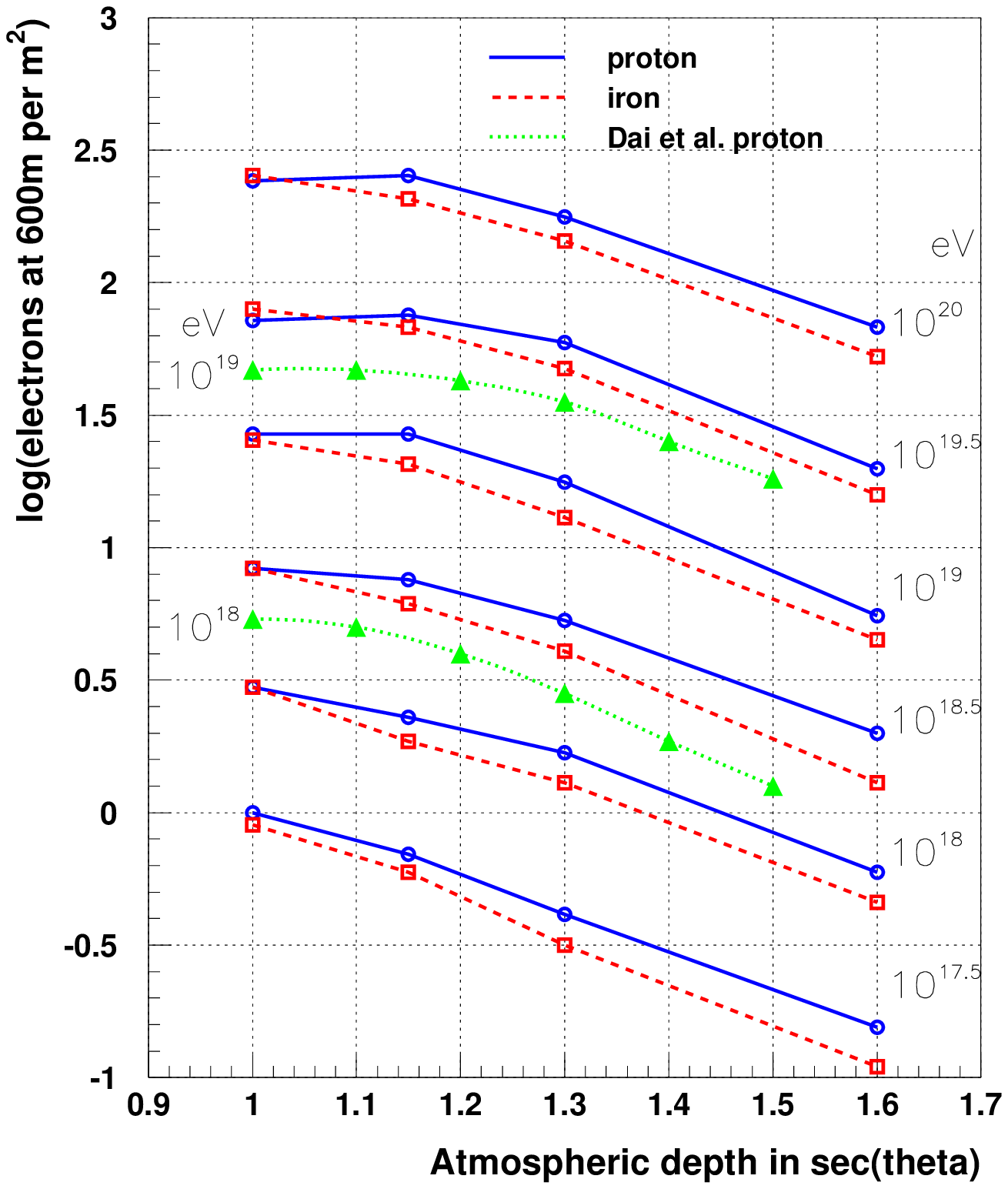,width=7cm,bbllx=1,bburx=377,bblly=16,bbury=460,clip=}
\hfill
\epsfig{file=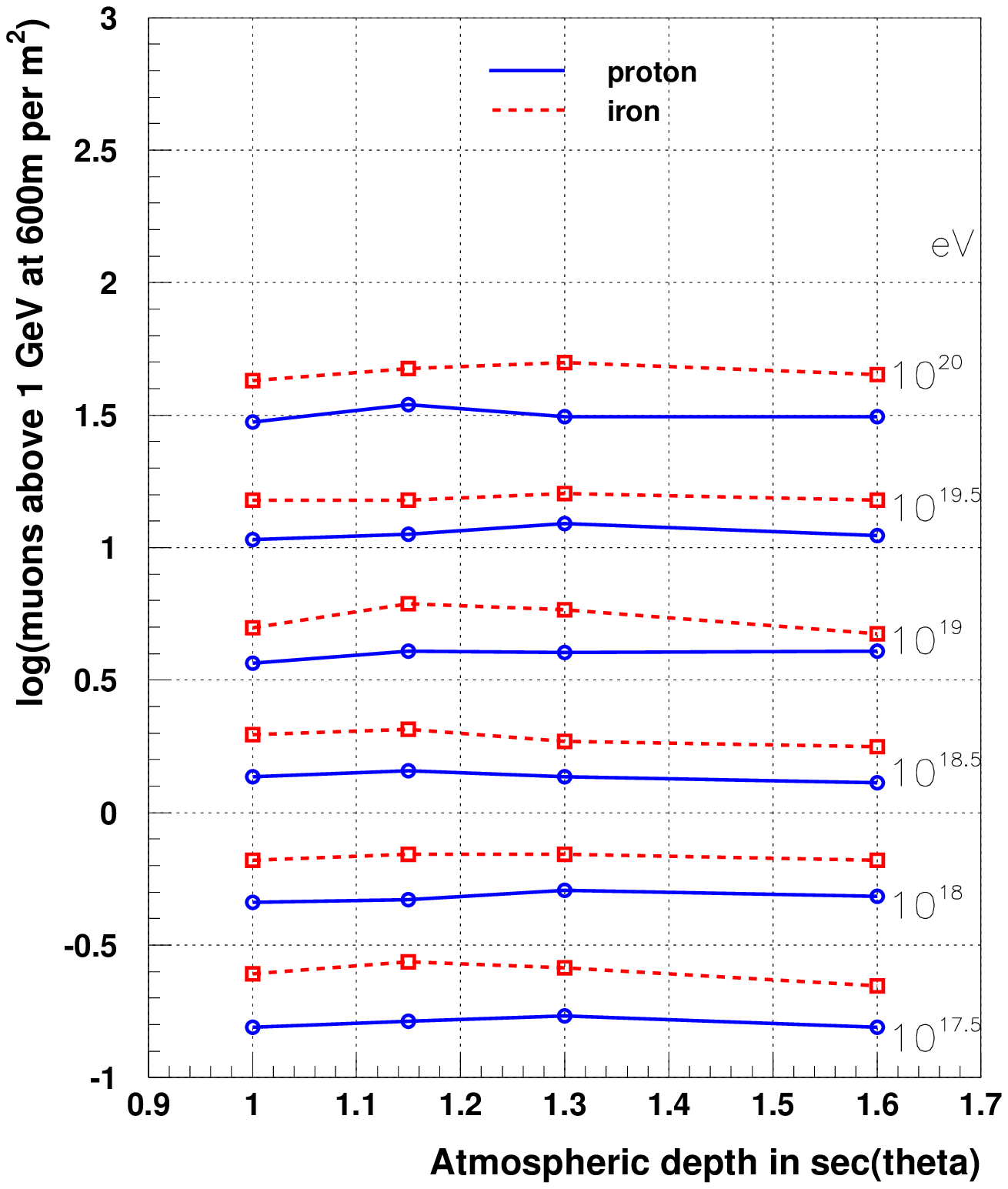,width=7cm,bbllx=1,bburx=377,bblly=16,bbury=460,clip=}
\end{center}
\caption{\small The variation of electron density at 600 m from the core
with zenith angle (left) and that of muon density above 1 GeV at 600 m
from the core (right).  Solid lines are from proton primaries and dotted
are from iron (QGSJET model, thinning level 10$^{-6}$ and cutoff energy
of electrons are 1.0 MeV).  Triangles connected with dotted lines are
from Dai {\it et al}. [6] for primary protons of 10$^{18.0}$ eV and
10$^{19.0}$ eV, respectively.}
\label{equi_e_mu600}
\end{figure}

To examine the difference, the zenith angle variation of the electron
density ($>1$ MeV) at 600 m from the core and the muon density ($>1$
GeV) at 600 m from the core are plotted in Fig. \ref{equi_e_mu600}.  The
irregularities in the curves may be due to the statistical fluctuations
of the limited number of simulated showers at each point and fitting
errors to derive the density at 600 m.  Within these uncertainties the
{\em attenuation} with zenith angle by CORSIKA and that by Dai
{\it et al.} agree with each other as in the left figure of
Fig. \ref{equi_e_mu600}.  The difference of absolute values is partly
due to the difference of cutoff energies as described before.  Since the
number of muons does not change with zenith angle and the number of
muons in iron initiated showers are larger than that in proton showers
as shown in the right figure of Fig. \ref{equi_e_mu600}, the attenuation
of $\rho_{ch}$(600), which consists of electrons above 1 MeV and muons
above 10 MeV, does not depend on primary energy and composition.

As is described before, $S_0$(600) must be evaluated as
$\rho_{sc}$(600), energy deposit in a scintillator in units of V$_{pw}$,
at various zenith angles taking into account the increase of
scintillator thickness with zenith angle.  It is found that the
attenuation of electromagnetic density with zenith angle is compensated
with muon density and hence the attenuation of $\rho_{sc}$(600) with
zenith angle is similar for proton and iron primaries in case of the
QGSJET model.  Therefore the muon density relative to the
electromagnetic density, which is sensitive to the hadronic model, is an
important quantity to compare with the experiment.  Since the AGASA data
have been increased considerably at the highest energy region compared
to the one used in the Fig. \ref{equi_sc600}, it may be better to await
further comparisons for the publication of new experimental results.

\section{Conclusion}
An interpretation of AGASA data has been made by comparing them with
results simulated with CORSIKA.  Ten air showers for each of the primary
energies 10$^{17.5}$ eV, 10$^{18.0}$ eV 10$^{18.5}$ eV, 10$^{19.0}$ eV
10$^{19.5}$ eV and 10$^{20.0}$ eV are simulated under each combination
of the QGSJET or SIBYLL hadronic models and proton or iron primaries.
General features of the electromagnetic component and low energy muons
observed by AGASA can be well reproduced by CORSIKA.  There are some
discrepancies which must be studied in more detail to improve the
agreement between the experiment and simulation.  In the following some
results implicated by the simulation related to the AGASA experiment are
summarized.

The form of the lateral distribution of charged particles agrees well
with the experimental one between a few hundred m and 2000 m from the
core, irrespective of hadronic interaction model or composition and it
does not depend on the primary energy simulated.  Though the shape of
the muon lateral distribution fits also the experiment, absolute values
change with the hadronic interaction model and primary composition.

If we evaluate the density measured by scintillator of 5 cm thickness as
$S_0$(600) by taking into account similar conditions as in the
experiment, the conversion relation from $S_0$(600) to the primary
energy can be written as
\begin{displaymath} 
  E [\mbox{eV}] = 2.15 \times 10^{17} S_0(600)^{1.015}.
\end{displaymath}  
This means that the energy assignment of the AGASA experiment is shifted
to higher energies, if we estimate it using CORSIKA.

The slope of $\rho_{\mu}$ vs. $S_0$(600) relation in the experiment is
flatter than that in simulations of any hadronic model and primary
composition.  The situation may not be changed even by taking into
account the primary energy spectrum and fluctuations of $\rho_{\mu}$ and
$S_0$(600).  Since the slope seems to be constant in a wide primary
energy range, we need to study this relation over a wide energy range.
Otherwise the composition may be interpreted as heavier than iron or
lighter than proton outside the narrow investigated energy region.

There is a disagreement of the attenuation length determined at AGASA
and that by CORSIKA simulation, even if we take into account the
particle incident angle to the scintillator.  If we take into account
the experimental error in zenith angle determination and $S_0$(600)
determination, the disagreement increases.  Since the experimental data
used in the present analysis is still preliminary, we better wait for
the publication of new experimental values for further discussions.

\section*{Acknowledgement}
The simulations have been performed by the facilities of the KASCADE
group of Institut f\"{u}r Kernphysik III, Forschungszentrum Karlsruhe
and the AGASA data published so far are used in this analysis.  We
acknowledge the kind cooperation of the KASCADE and the AGASA
collaborators and Profs. G. Schatz, H. Rebel and A.A. Watson for
valuable comments and suggestions.

M.N. wishes to thank the Alexander von Humboldt Stiftung and
Prof. G. Schatz for their support and warm hospitality during his stay
at Karlsruhe.  He also wishes to acknowledge all members of the KASCADE
group, especially Drs. A. Haungs and K. Bekk, who helped him in
performing this study efficiently.  He is also grateful to T. Kutter of
University of Chicago for his information before the publication and
N. Sakaki of ICRR, University of Tokyo who helped him to perform the
analysis.


\begin{thebibliography}{900}
\bibitem {grei66}
K. Greisen,  Phys. Rev. Lett. {\bf 16} (1966) 748;
Z.T. Zatsepin and V.A. Kuzmin, Pisma Zh. Eksp. Teor. Fiz. {\bf 4} 
 (1966) 144.
\bibitem {bird95}
D. Bird {\it et al}.,  Astrophys. J.  {\bf 441} (1995) 144.
\bibitem {hayashida94}
N. Hayashida  {\it et al}.,  Phys. Rev. Lett. {\bf 73}  (1994) 3491.
\bibitem {takeda98}
M. Takeda {\it et al}.,  Phys. Rev. Lett.  {\bf 81} (1998) 1163;
                   Astrophys. J. {\bf 522} (1999) 225.
\bibitem {hillas}
A.M. Hillas  {\it et al}.,  Proc. 12th ICRC (Hobart),  {\bf 3} (1971) 1001.
\bibitem {dai88}
H.Y. Dai {\it et al}.,  J. Phys. G: Nucl. Phys.  {\bf 14} (1988) 793.
\bibitem {kasahara}
K. Kasahara, S. Torii  and T. Yuda,   Proc. 16th ICRC (Kyoto)
 {\bf 13}  (1979) 70.
\bibitem {ding}
L.K. Ding {\it et al}.,  Proc. Int. Symp. on Cosmic Rays and Particle Physics
(ed. by T.Yuda, Inst. for Cosmic Ray Research, University of Tokyo) \
 (1984) 142.
\bibitem {nagano84}
M. Nagano  {\it et al}.,  J. Phys. Soc. Japan {\bf 53}  (1984) 1667.
\bibitem {law91}
M.A. Lawrence, R.J.O. Reid and A.A. Watson, J. Phys. G: Nucl. Phys. 
 {\bf 17} (1991) 733.
\bibitem {afa93}
B.N. Afanasiev {\it et al}., 
 Proc. Tokyo Workshop on Techniques for the Study of
the Extremely High Energy Cosmic Rays (ed. by M.Nagano,
Inst. for Cosmic Ray Research, University of Tokyo) \ (1993) 35. 
\bibitem {bird94}
D.J. Bird  {\it et al}.,  Astrophys. J. \ {\bf 424} (1994) 491.
\bibitem {sakaki}
M. Sakaki {\it et al}.,  Proc. 25th ICRC (Durban) \ {\bf 5}  (1997) 217. 
\bibitem {cronin97}
J.W. Cronin,  GAP-97-034 (Auger Technical Note) (1997).
\bibitem {kutter98}
T. Kutter,    GAP-98-048 (Auger Technical Note) (1998).
\bibitem {shibata65}
S. Shibata {\it et al}.,  Proc. 9th ICRC (London) {\bf 2} (1965) 672.
\bibitem {dawson98}
B. Dawson, R. Meyhandan and K.R. Simpson, 
 Astroparticle Phys. \ {\bf 9}  (1998) 331. 
\bibitem {gaisser93}
T.K. Gaisser {\it et al}.,  Phys. Rev. D \ {\bf 47} (1993) 1919.
\bibitem {nagano92}
M. Nagano {\it et al}.,  J. Phys. G: Nucl. Part. Phys. \ {\bf 18} (1992) 423.
\bibitem {yoshida94}
S. Yoshida {\it et al}., J. Phys. G: Nucl. Phys. \ {\bf 20} (1994) 651.
\bibitem {heck98a}
D. Heck {\it et al}.,  FZKA6019 (Forschungszentrum Karlsruhe) (1998).
\bibitem {heck98b}
D. Heck and J. Knapp,  FZKA6097 (Forschungszentrum Karlsruhe) (1998).
\bibitem {chiba92}
N. Chiba {\it et al}.,  Nucl. Instrum. and Methods \ {\bf A311}  (1992) 338.
\bibitem {ohoka96}
H. Ohoka {\it et al}.,  Nucl. Instrum. and Methods \ {\bf A385}  (1996) 268.
\bibitem {sakaki98}
N. Sakaki {\it et al}., Proc. 26th ICRC (Salt Lake City) \ {\bf 1} (1999) 361.
\bibitem {hayashida95}
N. Hayashida {\it et al}.,  J. Phys. G: Nucl. Phys. \ {\bf 21}  (1995) 1101.
\bibitem {doi95}
T. Doi  et al.,  Proc. 24th ICRC (Rome) \ {\bf 2}  (1995) 764.
\bibitem {matsu85}
Y. Matsubara {\it et al}., Proc. 19th ICRC  (La Jolla) \ {\bf 7} (1985) 119.
\bibitem {knapp}
J. Knapp, D.  Heck and G. Schatz,  FZKA5828 
(Forschungszentrum Karlsruhe) (1996).
\bibitem {kalmykov93}
N.N. Kalmykov and S.S. Ostapchenko,  Yad. Fiz \ {\bf 56} (1993) 105.
\bibitem {fletcher94}
R.S. Fletcher {\it et al}.,  Phys. Rev. D \ {\bf 50} (1994) 5710.
\bibitem {greisen56}
K. Greisen,  Prog. Cosmic Ray Physics, 3rd. (ed. by J.G. Wilson) (1956) p.27.
\bibitem {akeno}
Akeno Internal Manual, (1980) unpublished.
\bibitem {teshima86}
M. Teshima {\it et al}., J. Phys. G: Nucl. Phys. \ {\bf 12} (1995) 1097.
\bibitem {barnett96}
R.M. Barnett  {\it et al}.,  Phys. Rev. D \ {\bf 54}  (1996) 1.
\bibitem {leibrock98}
H. Leibrock, L. Haungs and H. Rebel,  unpublished report of  
Forschungszentrum Karlsruhe  (1998).
\bibitem {greisen60}
K. Greisen,  Ann. Rev. Nucl. Sci. {\bf 10} (1960) p.63.
\bibitem {hayashida97}
N. Hayashida {\it et al}.,  Proc. 25th ICRC (Durban) \ {\bf 4} (1997) 145.
\bibitem {murzin65}
V.S. Murzin,   Proc. 9th ICRC  (London) \ {\bf 2} (1965) 872.
\end{thebibliography}
\end{document}